\documentclass[12pt,a4paper]{iopart} 
\usepackage{iopams}  

\usepackage{amssymb}
\usepackage[mathscr]{eucal}
\usepackage{upgreek} 
\usepackage{array}
\usepackage{subfig}
\usepackage{graphicx}
\usepackage{bm}
\usepackage[figuresright]{rotating}

\usepackage[abs]{overpic}

\newcommand{\eqref}[1]{(\ref{#1})}
\newcommand{\figref}[1]{Fig.~\ref{#1}}
\newcommand{\Figref}[1]{Fig.~\ref{#1}}
\newcommand{\secref}[1]{section~\ref{#1}}
\newcommand{\Secref}[1]{Section~\ref{#1}}

\usepackage{afterpage}

 \usepackage{cite} 

\def\vec#1{\ensuremath{\mathchoice
                     {\mbox{\boldmath$\displaystyle\mathbf{#1}$}}
                     {\mbox{\boldmath$\textstyle\mathbf{#1}$}}
                     {\mbox{\boldmath$\scriptstyle\mathbf{#1}$}}
                     {\mbox{\boldmath$\scriptscriptstyle\mathbf{#1}$}}}}

\def\tens#1{\relax\ifmmode\mathsf{#1}\else\textsf{#1}\fi}

\renewcommand\div{{\rm div}}

\newcommand{\Bb}{{\boldsymbol{\mathnormal b}}}
\newcommand{\Bc}{{\boldsymbol{\mathnormal c}}}

\newcommand{\Bl}{{\boldsymbol{\mathnormal l}}}
\newcommand{\Bn}{{\boldsymbol{\mathnormal n}}}

\newcommand{\Bq}{{\boldsymbol{\mathnormal q}}}
\newcommand{\Br}{{\boldsymbol{\mathnormal r}}}

\newcommand{\BA}{\vec{A}}

\newcommand{\BQ}{\vec{Q}}

\newcommand{\dBI}{\ensuremath{\totdiff\Bl}} 

\newcommand{\subscr  }[1]{\ensuremath{{}_{\rm #1}}}

\newcommand{\Balpha} {\ensuremath{\boldsymbol\alpha}}

\newcommand{\Bvarrho}{\ensuremath{\boldsymbol\varrho}}

\newcommand{\rhot}   {\ensuremath{           \rho }} 
\newcommand{\qt}     {\ensuremath{                q}}

\newcommand{\rhoe}   {\ensuremath{           \varrho \subscr{e}}}

\renewcommand{\div}{\textrm{div}}

\newcommand  {\half     }{{\textstyle\frac{1}{2}}}

\newcommand{\totdiff}{\textrm{d}}

\newcommand{\dell   }{\ensuremath{\totdiff{\ell}}}
\newcommand{\du     }{\ensuremath{\totdiff{u}}}

\newcommand{\dx     }{\ensuremath{\totdiff{x}}}
\newcommand{\dy}{\ensuremath{\mathrm{d}y}}
\newcommand{\dz}{\ensuremath{\mathrm{d}z}}

\newcommand{\dBc    }{\ensuremath{\totdiff\Bc}}

\newcommand{\ddBc    }{\ensuremath{\totdiff^2\Bc}} 
\newcommand{\dJ}{\ensuremath{\totdiff J}} 

\usepackage[usenames, dvipsnames]{color}
\usepackage[normalem]{ulem} 

\newcommand \Insert [1] {\textcolor{red}{#1}}
\newcommand{\Delete} [1]{\bgroup\noindent\textcolor{red}{\xout{#1}}\egroup\ignorespacesafterend}
\newcommand{\SetTextColorRed}{\color{red}}
\newcommand{\SetTextColorBlack}{\color{black}}

\renewcommand{\SetTextColorRed}{\color{black}}
\renewcommand \Insert [1] {#1}
\renewcommand{\Delete} [1]{}

\begin{document}

\title[Microstructural Comparison of the Kinematics of DDD and CDD]{Microstructural Comparison of the Kinematics of Discrete and Continuum Dislocations Models.}

\author{Stefan Sandfeld$^1$ and Giacomo Po$^2$}
\address{$^1$Institute of Materials Simulation, University of Erlangen-N\"urnberg, Dr.-Mack-Stra{\ss}e 77, 90762 F\"urth, Germany}
\address{$^2$Mechanical and Aerospace Engineering Department, University of California Los Angeles, Los Angeles, CA, 90095, USA}
\ead{stefan.sandfeld@fau.de}

\begin{abstract}
The  Continuum Dislocation Dynamics (CDD) theory and the Discrete Dislocation Dynamics (DDD) method are compared based on concise mathematical formulations of the coarse graining of discrete data. A numerical tool for converting from a discrete to a continuum representation of a given dislocation configuration is developed, which allows to directly compare both simulation approaches based on continuum quantities (e.g. scalar density, geometrically necessary densities, mean curvature). Investigating the evolution of selected dislocation configurations \Insert{within analytically given velocity fields} for both DDD and CDD reveals that CDD contains a surprising number of important microstructural details.
\end{abstract}

\submitto{\MSMSE}


\section{Introduction}
\label{sec:intro}

Predicting and understanding the collective evolution of dislocation ensembles together with the resulting mechanical properties of crystalline materials is a long-standing goal of microstructure-based plasticity. From a computational perspective, reducing dislocation mechanics to a system of partial differential equations amenable to solution by standard continuum computational methods would be highly desirable. However, the nature of those curved, line-like defects makes the development of a continuum theory of dislocations an extremely difficult challenge \cite{Cottrell:2002vq}, which today is still not fully mastered. 
One way of enriching macroscopic continuum descriptions with additional details of smaller length scales (e.g. with details of dislocations) are concurrent as well as hierarchical multi-scale approaches for bridging the atomistic to the continuum domains \cite{Liu:2004fz,Curtin:2003tn,Zbib:2009tm}. These methods aim at establishing a direct link between atomistic details of dislocations and  macroscopic mechanical properties of materials. The level of detail in these approaches is very high, and length scales can be reached that are not accessible to one-scale methods as e.g. atomistic simulations. Multi-scale approaches, however, are still computationally very expensive and the information transfer between methods is numerically challenging.
An alternative description of dislocation ensembles is utilized in mesoscopic methods such as the discrete dislocation dynamics method (DDD) \cite{Lepinoux:1987tj,Ghoniem:1988wu,Gulluoglu:1989to,Kubin:1992wi,Schwarz:1997vo,Zbib:1998ub,Ghoniem:2000td,Weygand:2002tq,Bulatov:2004td,Po:2014en,Po:2014tc}. In DDD the atomic scale is not resolved, and instead dislocations are represented as discrete, linear defects interacting according to the elastic theory of dislocations \cite{Burgers:1939ui,Peach:1950tm,Friedel:1967voa,Nabarro:1980uk,Hirth:1992ub}. Because each dislocation is resolved individually, DDD is able to accurately describe the motion and interaction of dislocations in a very detailed manner. However, the computational cost of DDD scales with the number of dislocations considered and is therefore computationally expensive when it comes to large numbers of interacting dislocations (\emph{large} referring e.g. to a density of ${\geq}10^{13}{\rm m}^{-2}$ in a volume of ${\geq}(10\upmu{\rm m})^3$).
To overcome these limitations, one might seek continuous density-based descriptions as alternative to the representation of dislocations as discrete objects. Reducing dislocation microstructure to distinct ``types" of e.g. geometrically necessary and statistically stored dislocations (GNDs and SSDs) allows strain gradient based continuum methods \cite{Fleck1994_ActaMetallMater_p475,Gurtin2002_JMPS50_p5,Gao2003_ScrMater_p113} to become independent of the number of interacting dislocations. These approaches result in a substantial gain in terms of tractable length and time scales which makes them interesting for engineering applications. The main drawback is that only few, strongly averaged aspects of systems of dislocations are preserved, and e.g. fluxes of dislocations cannot be accounted for at all.
Recently emerging continuum models of dislocation dynamics (CDD) bridge the gap between these last two approaches by combining the efficiency of a continuum theory with the physically sound basis of a dislocation based description: 
\Insert{'Kr\"oner-Nye tensor'-related models \cite{Kroner1958,acharya01,Taupin2013370} are based on fluxes of GNDs,} and continuum screw-edge representations are a relatively coarse but efficient way of representing dislocation loops \cite{Arsenlis2004_JMPS52,Zaiser2006_ScriptaMater54, Reuber2014_ActaMater71}, while models that consider densities of curved lines contain considerably more information \cite{El-Azab2000_PhysRevB61,Sedlacek2003_PhilMag,Hochrainer2007_PhilMag,Xiang2009728,Sandfeld2010_JMR, Hochrainer2014_JMPS} and recently even have been coupled to gradient plasticity models \cite{Wulfinghoff2015_IJP}. Among others, applications with high densities and/or high accumulated plastic strains particularly benefit from continuum dislocation dynamics models (e.g. \Insert{\cite{Chen2013_IJP,Cheng2014_PhilMag,Sandfeld2015_MRS, Xia2015_MSMSE23,Sandfeld2015_MSMSE}}).

Because in a CDD-type continuum framework details of individual dislocations are accounted for in an average sense, one may question the physical validity and the predictive capability of such an approach. In principle it should be possible to benchmark CDD  theories through comparison with averaged data extracted from DDD simulations. Up to date, contributions into this direction have been presented by A. El-Azab and co-workers \cite{El-Azab2007_PhilMag_p1201,Deng2010_PhilMag90}, who investigated ensemble averages and statistical properties of dislocation microstructure e.g. in terms of dislocation density and orientation distributions. The objective of the present work is to compare CDD and DDD directly by analyzing dislocation microstructures that develop in time from equivalent initial conditions. Although such a detailed comparison is essential to validate  assumptions used to derive the particular CDD formulation, no previous attempts into this direction exist (to the best of the authors' knowledge).
One of the reasons for the lack of such direct comparisons may reside in the formulation of a CDD theory itself. In order to be comparable to DDD simulations in an average sense a CDD framework must fulfill a number of criteria: (i) its density measures must be derived by well defined statistical averaging steps from systems of discrete dislocations (i.e. in a bottom-up approach); (ii) its evolution equations should map an initial set of density measures onto a set of temporally evolved densities; (iii) it must be kinematically and (iv) dynamically consistent. \Insert{\begin{itemize}
	\item \emph{Kinematic consistency} describes the ability to evolve dislocations as curved and connected lines with the concomitant line length change during motion in a \emph{given} velocity field. 
	\item \emph{Dynamic consistency}, on the other hand, describes the ability to predict the actual evolution of a dislocation system under externally applied mechanical load as well as mutual interactions between dislocations. 
\end{itemize}}
\Insert{It is generally not recognized in the literature that } kinematic consistency is a necessary prerequisite for dynamic consistency and considers only geometrical aspects of averaged systems of dislocations. While the problem of dynamic consistency is to date still far from being solved, the recently developed (so-called 'simplified' or 'integrated') CDD theory by T. Hochrainer and co-workers \cite{Hochrainer2009_ICNAAM,Sandfeld2010_JMR,Hochrainer2014_JMPS,Hochrainer2015_PhilMag} was rigorously derived to solve the problem of kinematic consistency and has already demonstrated its applicability in a number of problems \cite{Sandfeld2010_JMR,Hochrainer2014_JMPS}, including  dislocation patterning \cite{Sandfeld2015_MSMSE}. \Insert{\emph{The goal of this work is to establish a methodology for a direct and detailed kinematic comparison between DDD and CDD models  and to demonstrate the accuracy of CDD through a number of benchmark problems}.} \\

The outline of this paper is as follows: 
in \secref{sec:coarsegrain} we give a brief overview of the general concept for validating continuum dislocation microstructure with discrete data used in this paper. 
In \secref{sec:averaging} we introduce the mathematical foundations for the geometrical representation of curved dislocations lines and for obtaining averaged continuum quantities in an analytical as well as in a numerical way. 
We then briefly introduce the evolution equations for CDD and how CDD data can be numerically obtained from DDD in \secref{sec:evoeqns}. \Secref{sec:examples} investigates numerically the quality of the continuum model during time evolution of DDD and CDD microstructures for three different benchmark systems. 
Finally, in \secref{sec:fitting} we demonstrate how the DDD model can be used for ``data mining" in order to estimate the quality of one of the assumptions on which the CDD version used in this paper is based.

\section{Outline of the CDD-DDD Validation Approach}
\label{sec:coarsegrain}
\begin{figure}[ht]
\centering
\includegraphics[width=\textwidth]{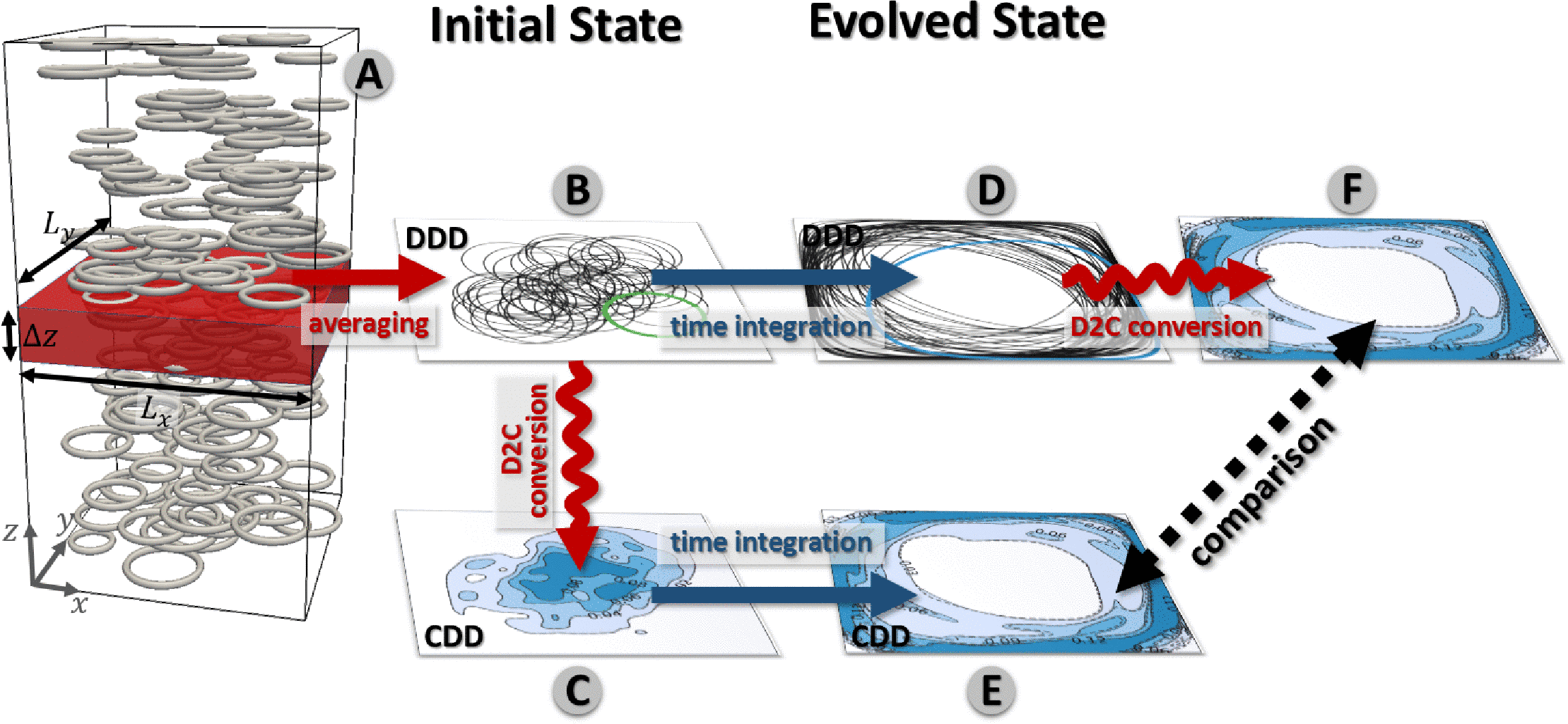}
\caption{\label{fig:sketch}
	Sketch of the system (A), coarse graining steps and time evolution. The continuous field (C) is obtained from the DDD configuration (B). Both systems then evolve independently. DDD and CDD result in (D) and (E), respectively, and can then directly be compared by converting (D) into continuous fields (F).}
\end{figure}

The fundamental idea for validating the microstructure evolution within CDD is sketched in \figref{fig:sketch}. 
Consider an ensemble of dislocation loops with the same Burgers vector $\bm b$, occupying a domain of size $L_x\times L_y\times \triangle z$ (\figref{fig:sketch}A), where the slip system normal points in $z$ direction. Each dislocation is located on a different slip plane and moves - driven by an external stress field - by glide only; no annihilation, cross slip etc are considered. Averaging over the height of a sub-volume $\triangle z$ we obtain \figref{fig:sketch}B, which serves as initial dislocation microstructure for our investigations. In the following sections a systematic method for extracting average fields (e.g. dislocation densities, Nye tensor, curvature, etc) from systems of discrete loops is developed. This \emph{discrete-to-continuum} conversion method (\emph{D2C}) is applied to the DDD initial configuration (\figref{fig:sketch}B), from which we obtain average fields (\figref{fig:sketch}C). Those serve as initial values for the CDD model. 
CDD initial structure (\figref{fig:sketch}C) is evolved by time integration of the CDD evolution equations, while the initial configuration of discrete loops is \textit{independently} evolved in time by the DDD equation of motion. This results in the DDD and CDD microstructures as shown in \figref{fig:sketch}D and E. At any subsequent point of time, relevant state quantities of CDD, \figref{fig:sketch}E, can be compared to corresponding DDD quantities as obtained from the \emph{D2C} conversion, \figref{fig:sketch}F.

\section{Conversion of discrete dislocation lines into continuous fields (D2C)}
\label{sec:averaging}
In what follows the discrete-to-continuum (\emph{D2C}) approach is described which is used to systematically average the geometric properties of discrete dislocation lines to obtain CDD state quantities. To differentiate between variables for discrete and continuous measures the superscript 'd' is used for discrete quantities. The superscript is dropped for a continuous quantity or when the context is clear. This section starts by introducing the geometrical representation of discrete lines and their approximation through splines, followed by a formal density definition. Subsequently, mathematical averaging operators are introduced and concisely derived in a discretized form which is suitable for numerical implementation.


\subsection{Discrete dislocation lines and their geometrical approximation} 

From a geometrical viewpoint, a dislocation line is an oriented curve $\Bc(\ell)$ which can be parameterized by its arc-length $\ell$. The local unit tangent vectors are the first derivative of $\Bc$
\begin{equation}
\Bl^{\rm d}(\ell)=\frac{\dBc}{\dell}\, , \label{eq:unit_tangent}
\end{equation}
 while the curvature vector is given by the change of orientation per unit arc-length, 
\begin{equation}
\bm k^{\rm d} =\frac{\dBI^{\rm d}}{\dell}=\frac{\ddBc}{\dell^2}\, .
\end{equation}
The scalar curvature is the reciprocal of the local radius of curvature $R$  of the dislocation line: 
\begin{equation}
k^{\rm d} =\|\bm k^{\rm d}\|=\frac{1}{R} .
\end{equation}
%
%
Although parameterization by arc-length  is the most natural choice for defining geometrical properties of a curve, in general the parameter $\ell$ is only available \emph{after} the curve  has been constructed. Thus, for practical purposes, it is often necessary to prescribe a curve through a parameter $u$, say $u\in[0,1]$. In this case unit tangent and curvature become
\begin{equation}
\Bl^{\rm d}(u)=\frac{1}{J}\frac{\dBc}{\du}
\qquad \textrm{and}\qquad
\bm k^{\rm d}(u)=\frac{1}{J^2}\frac{\ddBc}{\du^2}-\frac{1}{J^3}\frac{\dJ}{\du}\frac{\dBc}{\du}
\end{equation}
where the scalar $J$ and its derivative are given by
\begin{eqnarray}
 J(u)&=&\frac{\dell}{\du}=\left\|\frac{\dBc}{\du}\right\|
 \label{J}
\qquad\textrm{and}\qquad
\frac{\totdiff J}{\du}=\frac{1}{J}\frac{\dBc}{\du}\cdot\frac{\ddBc}{\du^2}\, .
\end{eqnarray} 
In this work dislocation lines are discretized into segments, each of which is represented by a cubic Hermite spline of the form
\begin{equation}
\begin{array}{llll}
\bm c(u)
\end{array}
=
\left[
\begin{array}{llll}
1&u&u^2&u^3
\end{array}
\right]
\left[
\begin{array}{rrrr}
1&0&0&0\\
0&\gamma&0&0\\
-3&-2\gamma&3&-\gamma\\
2&\gamma&-2&\gamma
\end{array}\right]
\left[
\begin{array}{l}
\bm P_0\\
\bm T_0\\
\bm P_{1}\\
\bm T_{1}
\end{array}
\right],
\label{HSFC}
\end{equation}
where $(\bm P_0,\bm T_0)$ are position and parametric tangent vectors of the first knot of the segment ($u=0$),  $(\bm P_1,\bm T_1)$ are the corresponding quantities for the second knot of the segment ($u=1$), $\gamma=\|\bm P_1-\bm P_0\|^\alpha$, and $\alpha$ is a tension parameter\footnote{Parameterization corresponding to $\alpha=0$, $\alpha=0.5$, and $\alpha=1$ are called uniform, centripetal, and  chordal, respectively.}. Cubic splines are used instead of linear segments in order to have non-vanishing curvature along each segment.  This type of representation is adopted in the so-called Parametric Dislocation Dynamics method (PDD) \cite{Ghoniem:2000td}. The numerical studies below are based on the existing implementation \cite{Po:2014en,MODEL:2013}.

\subsection{Averaging dislocation fields}
\label{sec:maths}

As an auxiliary functional for averaging geometrical properties of discrete dislocations, we introduce the Dirac delta-operator on a dislocation line $\bm c$ as
\begin{eqnarray}
  \delta_\Bc\left([ \bullet ], \Br\right) = \int\limits^{L_\Bc}_{0} \delta(\Bc(\ell)-\Br) \bullet {\rm d}\ell,
  \label{directDelta}
\end{eqnarray}
where the symbol '$\bullet$' is a placeholder for an arbitrary line property described by a scalar, vectorial or tensorial function of $\Bc$. Using \eqref{directDelta}, for each discrete line $\bm c$ we can define the scalar dislocation density $\rho_\Bc$, the line direction density (or \emph{vector of signed GND densities}) $\bm\varrho_\Bc$, the Kr\"oner-Nye tensor $\bm\alpha_\Bc$ \cite{Nye:1953vn}, and the curvature density vector $\Bq_\Bc$  as\footnote{In the following we assume only one slip system, i.e. all dislocations have the same Burgers vector $\Bb$. From this case we can then construct the multi-slip system situation by appropriate superposition (e.g. summation of densities).}
\begin{eqnarray}
 \rho_\Bc (\bm r)= \delta_\Bc\left([ 1],\Br\right)= \int\limits^{L_\Bc}_{0} \delta(\Bc(\ell)-\Br) {\rm d}\ell
\label{eq:rhoc}
\end{eqnarray}
\begin{eqnarray}
  \bm\varrho_\Bc (\bm r) = \delta_\Bc\left(\left[\frac{{\rm d}\Bc}{{\rm d}\ell}\right],\Br\right)= \int\limits^{L_\Bc}_{0} \delta(\Bc(\ell)-\Br)  \frac{{\rm d}\Bc}{{\rm d}\ell}\, {\rm d} \ell
  \end{eqnarray}
\begin{eqnarray}
  \Balpha_\Bc (\bm r) =\delta_\Bc\left(\left[\frac{{\rm d}\Bc}{{\rm d}\ell}\otimes\Bb\right],\Br\right)= \int\limits^{L_\Bc}_{0} \delta(\Bc(\ell)-\Br)  \frac{{\rm d}\Bc}{{\rm d}\ell} \otimes \Bb\, {\rm d} \ell
\end{eqnarray}
\begin{eqnarray}
\Bq_\Bc (\bm r) = \delta_\Bc\left(\left[\frac{\ddBc}{{\rm d}\ell^2}\right],\Br\right)= \int\limits^{L_\Bc}_{0} \delta(\Bc(\ell)-\Br)\,  \frac{\ddBc}{\dell^2}
\, {\rm d} \ell.
\end{eqnarray}
The later on used scalar, signed curvature density $q_\Bc$ can be obtained by projecting the curvature vector on the outwards pointing normal
\begin{eqnarray}
q_\Bc (\bm r) = 
 \int\limits^{L_\Bc}_{0} \delta(\Bc(\ell)-\Br)\,  \frac{\ddBc}{\dell^2}\cdot\left(\frac{{\rm d}\Bc}{{\rm d}\ell}\times{\Bn}\right) \, {\rm d} \ell,
\label{eq:qc}
\end{eqnarray}
where $\Bn$ is the slip plane normal. These discrete measures are well suited for spatial (or ensemble) averaging over a number of lines, and we introduce an averaging operator for the volume $V_{\Br}$ of size $V$ as
\begin{eqnarray} \label{eq:average_op}
  \langle \bullet \rangle_{V,\Br} := (1/V)\int\limits_{V_{\Br}}\bullet\; \totdiff^3 r,
  \label{eq:avrg_op}
\end{eqnarray}
where the averaging volume is centered around $\Br$. Application of \eqref{eq:avrg_op} to \eqref{eq:rhoc}-\eqref{eq:qc} leads to the definition of the corresponding average fields: 
\begin{eqnarray}
\rhot(\bm r) = \left\langle \sum_\Bc \rho_\Bc \right\rangle_{V,\Br}=\frac{1}{V}\sum_\Bc \int_{\mathcal{L}_{\Bc}^{V,\bm r}}\, d\ell 
   \label{eq:rho_stat_av}
\end{eqnarray}
\begin{eqnarray} \label{eq:varrho}
  \Bvarrho(\bm r) = \left\langle \sum_\Bc   \Bvarrho_\Bc \right\rangle=\frac{1}{V}\sum_\Bc \int_{\mathcal{L}_{\Bc}^{V,\bm r}} \frac{{\rm d}\Bc}{{\rm d}\ell}\, d\ell 
\end{eqnarray}
\begin{eqnarray}
  \Balpha(\bm r) = \left\langle \sum_\Bc \Balpha_\Bc \right\rangle=\frac{1}{V}\sum_\Bc \int_{\mathcal{L}_{\Bc}^{V,\bm r}} \frac{{\rm d}\Bc}{{\rm d}\ell} \otimes \Bb\, {\rm d} \ell
\end{eqnarray}
%
%
\begin{eqnarray}
\qt(\bm r) =  \left\langle \sum_\Bc q_\Bc\right\rangle=\frac{1}{V}\sum_\Bc \int_{\mathcal{L}_{\Bc}^{V,\bm r}}\,\frac{\ddBc}{\dell^2}\cdot\left(\frac{\dBc}{\dell}\times\Bn\right)
\, {\rm d} \ell
\label{eq:q_stat_av_signed}
\end{eqnarray}

Therein, $\mathcal{L}_{\Bc}^{V,\bm r}\subset \Bc$ denotes a section of line $\Bc$ contained inside the averaging volume $V$ which is centered at $\bm r$. Note, that upon volume integration of the total density \eqref{eq:rho_stat_av} the total dislocation line length contained within the volume is obtained. Analogously, integration of the signed curvature density \eqref{eq:q_stat_av_signed} yields the number of closed loops as multiple of $2\pi$. The vector of signed GND densities $\Bvarrho$ contains densities of positive and negative edge and screw dislocations as later used in the CDD evolution equation  \eqref{eq:dvarrhodt}. We remark that the (scalar) {geometrically necessary dislocation density} $\rho_{\rm G}$ is the norm of $\Bvarrho$, i.e. $\rho_{\rm G}\equiv\varrho=\|\Bvarrho\|$, and that the average line direction of the geometrically necessary dislocations  is given by the unit vector $\Bl_{\Bvarrho}=\Bvarrho/\|\Bvarrho\|$.

\subsection{Numerical discretization of the averaging scheme}
 
Numerically, the fields \eqref{eq:rho_stat_av}-\eqref{eq:q_stat_av_signed} can be discretized by replacing the spatial points $\bm r$ with a finite set of  points $\bm r_i$  which have the coordinates $(x_i,y_i,z_i)$. On a regular grid each point $\Br_i$ is then the center of a sub-volume $V_i=\triangle x \triangle y \triangle z$ which defines the averaging sub-domain $\Omega_i$:
\begin{eqnarray}
\Omega_i = \left[x_i-\half\triangle x, x_i+\half\triangle x\right] \times 
\ldots\times \left[z_i-\half\triangle z, z_i+\half\triangle x\right]
\end{eqnarray}
and determines the resolution of the continuum representation. As a consequence of this coarse graining, it is well possible that inside a sub-domain $\Omega_i$ line segments ``cross each other" without intersections: although these lines are located on different physical glide planes they can not be resolved separately after coarse graining\footnote[1]{The coarse graining resolution is the reason why short-range interactions have to be handled differently as compared to DDD. This, however, will not be discussed here.}. 
The discretized version of the scalar dislocation density \eqref{eq:rho_stat_av} is:
\begin{eqnarray} \label{eq:num_averagre}
\rho_i =\rho(\bm r_i)=\frac{1}{V_i}\sum_\Bc\left(\sum_{k,\, \bm c(u_k)\in \Omega_i} \!\!\! J_\Bc(u_k)w_k \right).
\label{eq:rho_i}
\end{eqnarray}
%
{In \eqref{eq:rho_i} we compute the line length contribution of each dislocation segment by numerical quadrature: 
	letting $u_k$ and $w_k$ be the abscissas and weights of the quadrature method of choice\footnote{%
		In this work we have used a uniform distribution of the quadrature points $i$ the interval $[0,1]$.} in the unit interval $[0,1]$, respectively,}
%
each quadrature point contributes a line length $J(u_k)w_k$ to the domain $\Omega_i$ which contains $\bm c(u_k)$. Conveniently, the quantity $J(u_k)$ can be directly computed for each dislocation segment from \eqref{J} and \eqref{HSFC}. Analogously, we define the discretized vector of signed GND densities, and the discretized signed curvature density as
\begin{eqnarray} \label{eq:num_averagre2}
\bm\varrho_i =\bm\varrho(\bm r_i)=\frac{1}{V_i}\sum_\Bc\left(\sum_{k,\, \bm c(u_k)\in \Omega_i} \frac{{\rm d}\Bc}{{\rm d}\ell}(u_k)\,J_\Bc(u_k)w_k \right)
\label{eq:varrho_i}
\end{eqnarray}
%
%
%
\begin{eqnarray}\label{eq:q_i}
   q_i =q(\bm r_i)=\frac{1}{V_i}\sum_\Bc\left(\sum_{k,\, \bm c(u_k)\in \Omega_i} \left[\frac{\ddBc}{\dell^2}\cdot\left(\frac{\dBc}{\dell}\times{\Bn}\right)\right]_{u_k} \!\!\!J_\Bc(u_k)w_k \right).
\end{eqnarray}

\section{Evolution equations for CDD and DDD}
 \label{sec:evoeqns}
 
\subsection{Continuum Dislocation Dynamics}
The Continuum Dislocation Dynamics theory provides conceptual steps to map a discrete dislocation system onto a set of continuous, density-like field variables together with the respective evolution equations governing their change in time. The original, higher-dimensional theory (hdCDD) was defined in a configuration space which added an additional dimension to the spatial domain \cite{Hochrainer2007_PhilMag,Sandfeld2010_PhilMag90}. hdCDD is very accurate and contains very detailed microstructure information \cite{Sandfeld2015_IJP, Sandfeld2015_MRS} but suffers from requiring a large number of computational degrees of freedoms. As a computationally favorable albeit somewhat less accurate theory a simplified version (CDD) is used, which is based on spatial evolution equations for the total dislocation density $\rhot$, a vector of geometrically necessary signed densities $\Bvarrho$ and the curvature density $\qt$ \cite{Hochrainer2014_JMPS,Monavari2014_MRSSP,Hochrainer2015_PhilMag}. We assume that the components of $\Bvarrho$ are oriented such that they represent the orientation of screw and edge dislocations: $\Bvarrho=[\varrho_{\rm s}, \varrho_{\rm e}]$. \Insert{The temporal evolution of these field quantities are -- in local slip system coordinates -- governed by the following set of equations:
%
%
%
\begin{eqnarray}
\label{eq:drhotdt}
\partial_t\rhot &=&-\nabla\cdot(v\Bvarrho^\perp)+v\qt\\
\label{eq:dvarrhodt}
\partial_t\Bvarrho &=& -\nabla\times(v\rhot\Bn)\\
\label{eq:dqtdt}
\partial_t\qt &=&-\nabla\cdot( -v\BQ^{(1)} + \BA^{(2)}\cdot \nabla v ),
\end{eqnarray}
}%
where $\Bn$ denotes the slip plane normal. The by $90^\circ$ rotated GND density vector is $\Bvarrho^\perp=[\varrho_{\rm e},-\varrho_{\rm s}]$, and we assume $\BQ^{(1)}=- \Bvarrho^{\perp} {\qt}/{\rhot}$ (see \cite{Monavari2014_MRSSP} for a discussion of this assumption). The evolution equations are mathematically closed by an expression for the tensor $\BA^{(2)}$ which here is obtained from a ``maximum entropy" approach \cite{Monavari2014_MRSSP}:
\begin{equation}\label{eq:ME_A2}
{\BA}^{(2)} = \frac{\rhot}{2}\left[ (1+\Phi) {\Bl}_{\varrho} \otimes {\Bl}_{\varrho} + (1-\Phi){\Bl}_{\varrho^\perp} \otimes {\Bl}_{\varrho^\perp}\right] 
\end{equation}
where ${\Bl}_{\varrho^\perp}$ is the unit vector perpendicular to ${\Bl}_{\varrho}$. As shown by Monavari et al. \cite{Monavari2014_MRSSP} $\Phi$ can be approximated as $\Phi \approx  (\varrho/\rhot)^2(1+(\varrho/\rhot)^4)/2$ and is a non-linear interpolation between $\varrho =0$ (isotropic dislocation arrangement) and $\varrho = \rhot$ (fully polarized dislocation arrangement).  In \secref{sec:fitting} we numerically investigate the validity of this approximation. The corresponding non-dimensionless scaling of the CDD equations \eqref{eq:drhotdt}-\eqref{eq:dqtdt} can be found in Appendix D of \cite{Zaiser2014_MSMSE22}.

In general, the velocity $v$ in the above equations needs to be specified in terms of a constitutive equation through which elastic interactions enter the model. This is a priori not part of CDD: e.g. long-range (or ``internal") interaction stresses have to be obtained from the additional solution of a dislocation eigenstrain problem \cite{Kroener1955_ZPhys, El-Azab2007_PhilMag_p1201, Sandfeld2013_MSMSE} while elastic short-range interactions have to be recovered in an alternative way based on CDD fields (see e.g. \cite{Groma1997_PhysRevB_p5807,Groma2003_ActaMater,Kooiman2015_JMPS78} for steps into this direction) due to the limited resolution of any continuous description. Within the present work we will not tackle this problem of ``dynamic closure" and assume for benchmark purposes velocity fields given as stationary analytical functions that do not depend on the dislocation state\footnote{Note that this assumption can be physical e.g. in the case of a high purity semiconductor single crystal where the Peierls barriers are high and dislocation densities are low such that the influence of dislocation interactions on their motion can be neglected. In general, of course, interactions cannot be neglected. }. 
In other words: we ignore all dislocation interactions and concentrate exclusively on the ``kinematics", that is, how curved and connected lines move and evolve in space and time without interactions.

\subsection{Regularization}
\label{sec:regularization}
Eqs.~\eqref{eq:rho_i}-\eqref{eq:q_i} can be used to extract coarse grained fields of geometrical data from a given configuration of discrete dislocations. However, the CDD evolution equations contain spatial derivatives which have to be approximated numerically and which thus require field data with a certain level of spatial smoothness. It is therefore helpful to replace the Dirac delta-function $\delta(\Br)$ in the convolution integral \eqref{directDelta} with a regularization function $G(\Br)$. To improve the numerical efficiency during computation of the convolution it is beneficial to make use of the sifting property of the Dirac function: we can always write the regularized function as $G=G*\delta$, where the symbol $*$ indicates convolution over three-dimensional space. Together with the fact that convolution and volume averaging commute, it follows that regularized fields can be obtained by convolution of \eqref{eq:rho_i}-\eqref{eq:q_i} with $G$. In our numerical implementation, we chose a Gaussian standard distribution as the regularization function: 
\begin{eqnarray} \label{eq:rho_gauss}
	G(\bm r)=           \frac{1}{s\sqrt{2\pi}}\exp\left(-\frac{\|\Br\|^2}{2s^2}\right),
\end{eqnarray}
where 
the standard deviation $s$ characterizes the width of the dislocation density distribution. For numerical reasons $G$ is approximated by a \emph{discrete} Gauss function $G^{\rm d}$ which is non-zero only at a finite number of discrete points $r_i$ and is normalized such that $\triangle x \triangle y \triangle z\sum_i G^{\rm d}(i) = \int_{x,y}G(r(x',y',z')) \dx'\,\dy'\,\dz'$ within a small numerical tolerance. We then compute the convolution of the coarse grained fields (eqns. \eqref{eq:rho_i}-\eqref{eq:q_i}) with the discrete Gauss function $G^{\rm d}$. We note, that in \eqref{eq:rho_gauss} the only free parameter is the standard deviation $s$ which subsequently will be chosen as the average dislocation spacing $\bar{s}$ (the mean separation between a dislocation and its nearest neighbors) inside the averaging volume (see also the discussion in \cite{Sandfeld2013_MSMSE}). \Insert{Note that choosing a much larger value for $s$ would destroy details of the dislocation structure to a large extend while if $s$ were very small this would result in a quasi-discrete rather than continuous density structure. This is further discussed and demonstrated in the appendix.} \Delete{As a rule of thumb, we find that microstructure evolution does not depend on the particular choice of $s$ within a range of approximately $[\bar{s}/3, 3\bar{s}]$.}

\section{Numerical experiments}
\label{sec:examples}
One of the key advantages of the CDD theory is the property of \emph{kinematic consistency}, i.e. the ability to represent flowing dislocations through their characteristic geometrical properties: a dislocation is a linear, curved line where each of its points moves perpendicular to the respective local line tangents.  \Insert{In the following benchmark systems we will directly compare the evolution of discrete and continuous dislocation microstructure in analytically given velocity fields. We start with 3 situations of increasing complexity with regards to boundary conditions}: dislocation loop distributions in a rectangular domain (i) with  open boundaries allowing out-flux of dislocations,  (ii) with impenetrable surfaces and (iii) with a circular obstacle (i.e. an internal boundary). All 3 systems will start from the same initial dislocation microstructure which makes it easy to compare the effects caused by  different external and internal boundaries.
\Insert{The 4th benchmark system is a periodic system with a statistically homogeneous random  distribution of dislocations and investigates how good fluctuation can be represented in a situation without the strong polarizing effect of boundaries.}

\subsection{System and initial values}
\label{sec:sys_and_inivals}
The following examples study a system as sketched in \figref{fig:sketch} where a 3-dimensional pillar with a square base is discretized into thin sub-volumes along the $z$ axis. We investigate one of such sub-volumes with size $L_x=L_y=2000b$ and height $\triangle z=75b$, where $b=0.256\,{\rm nm}$ is the norm of the Burgers vector. 
Initial values for DDD simulations \Insert{for Systems 1-3} are obtained by randomly distributing $N_{\rm d}=50$ circular, discrete dislocation loops with radius $R=75b$ inside the volume $V$ (note, that we only chose \emph{circular} loops for ease of implementation; loops could be arbitrarily shaped - as long as they are closed). Each loop is positioned on a different glide plane and thus loops do not intersect by construction;  for the first three systems their centers are distributed on a quadratic domain $0.5L_x\times 0.5L_y$  such that all loops are completely contained inside the computational domain. The average density in this volume is 
$\rho_0=4.8\times 10^{15} {\rm m}^{-2}$. 
$\triangle z$ is the coarse graining height such that the system can be numerically represented as a two-dimensional object -- a thin lamella in the $x-y$ plane (\figref{fig:sketch}A). Continuous field data is obtained by 'smearing-out' the coarse grained fields through the regularization  outlined in \secref{sec:regularization}. \Insert{Initial values for System 4 are obtained by distributing the loop centers on the whole quadratic domain  $L_x\times L_y$ while simultaneously taking care of the periodicity. 
}
For determining the standard deviation for the Gaussian distribution we estimate the mean dislocation spacing from the density $\rho_0$, 
which results in a mean dislocation spacing of $s=L_x/(2\sqrt{2\pi N_{\rm d}})\approx L_x/35$. For our examples this estimate of the standard deviation works well for all investigated systems and at all time steps \Insert{(further details on the choise of $s$ can be found in the appendix).} \Delete{, and the microstructure evolution is largely independent of $s$ (e.g. all values of $s\approx L_x/100\ldots L_x/10$ work fairly well).} Only when the mean dislocation spacing changes drastically -- in \secref{sec:cell} due to dislocation pile ups at grain boundaries -- we adjust this value locally during the post-processing step for obtaining reference DDD field data to the actual value of the dislocation spacing. 
\begin{figure}[htb]
	\centering
	\includegraphics[width=\textwidth]{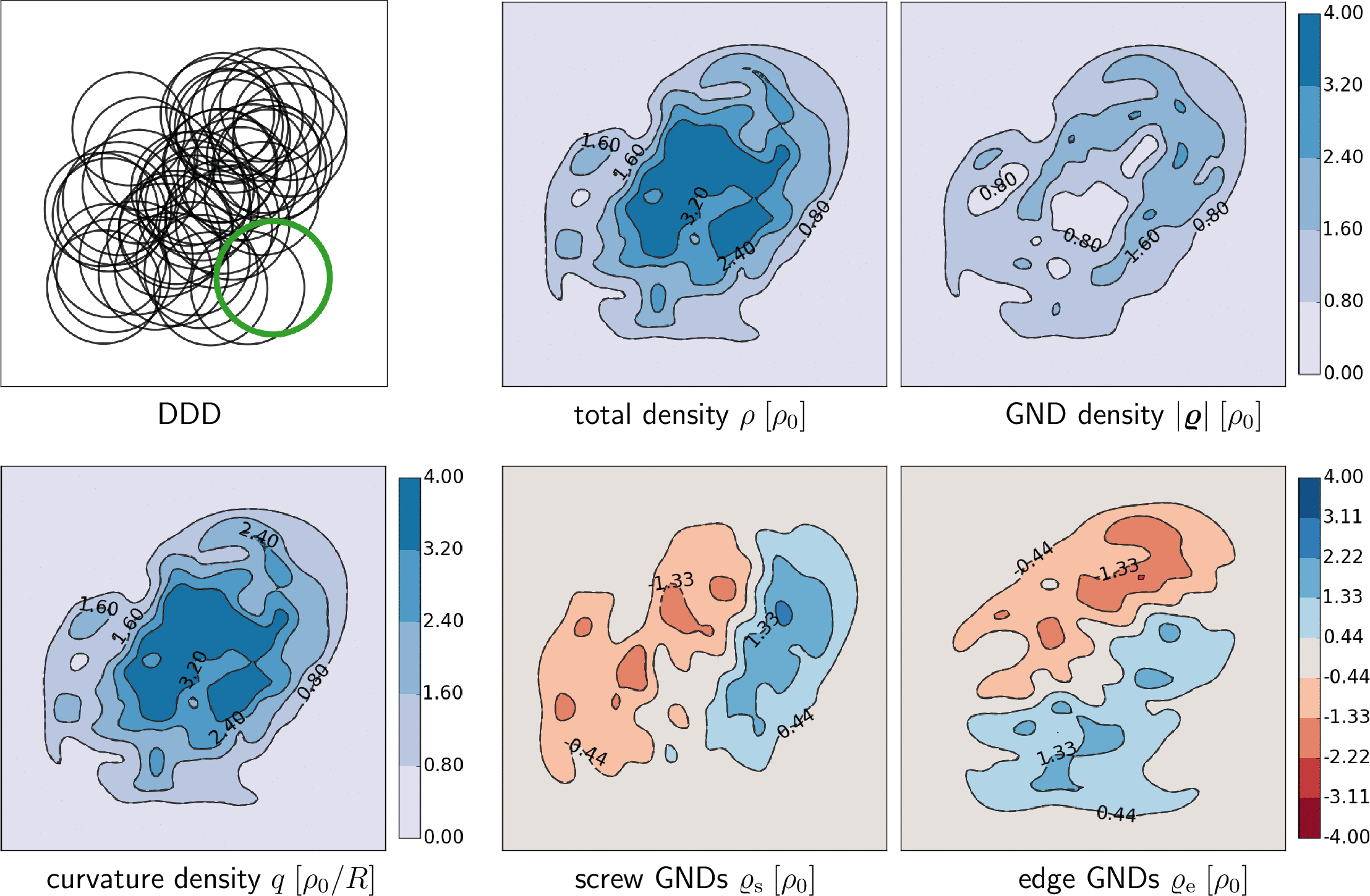}
	\caption{\label{fig:inivals} Initial values for the DDD simulation (upper left) from which initial values for the CDD field variables are obtained. }
\end{figure}
An example for DDD and resulting CDD initial structure that is used for all numerical examples is shown in \figref{fig:inivals}: the total dislocation density has a maximum in the center region where most of the loops overlap. At the same time nearly equal numbers of loop segments of all possible orientations are present in the center region which makes the dislocations 'statistically stored'. Hence, the GND density $|\Bvarrho|$ tends to zero. The curvature density $\qt$ has the same shape as the total density because initially all loops have the same radius and thus $\qt=\rhot/R$. \Insert{Although for each of the first 3 studies identical initial values are used, we note that the chosen CDD fields are not special or more suitable than any other set of statistically equivalent initial values.}

\subsection{Numerical methods}
For the numerical solution of the continuum evolution equations a Galerkin finite element scheme together with an implicit time integration scheme is used. To accurately represent derivatives of up to second order, Lagrangian shape functions with cubic polynomials were chosen. The finite element mesh consists of 30x30 elements and thus has an element size which is in the same range as $\triangle z$. When high density gradients at boundaries occur (System 2 and 3) we additionally refine elements in those regions. To increase numerical stability in particular in regions with high fluxes and where the density approaches zero a small viscous term was added to each evolution equation. E.g. in \eqref{eq:drhotdt} we replace the divergence term
$\nabla\cdot(v \Bvarrho)$ with $\nabla\cdot(v\Bvarrho  + \epsilon\nabla\Bvarrho)$, where $\epsilon$ is chosen sufficiently small such that the physical behavior of the equations is not affected while numerical oscillations are suppressed. The FEM simulations presented below all take less than two minutes computational time on a off-the-shelve computer with a 2.6 GHz dual core processor.

Computation of the evolution of plastic strain can be done for CDD based on the Orowan equation ($\partial\gamma_t=\rhot b v$) by time integration alongside with the CDD evolution equations \eqref{eq:drhotdt}-\eqref{eq:dqtdt}. To obtain the initial values for $\gamma$ from DDD one can compute $\gamma$ from the GND vector and the relation $\Bvarrho=-\frac{1}{b}\nabla\gamma$: geometrical construction of the swept area for each unit of GND density and subsequent superposition of these areas - followed by division by $\triangle z$ -  gives the plastic strain. Since the plastic strain is a derived quantity it suffices to show that main CDD field quantities match those from DDD and we do not explicitly show $\gamma$. 

The DDD model used the representation by chordal splines ($\alpha=1$); for further numerical details on the used Parametric Dislocation Dynamics method please refer to \cite{Po:2014en,MODEL:2013}.

\subsection{System 1: distribution of dislocation loops, open boundaries}
\begin{figure}[htb]
	\centering
	\includegraphics[width=\textwidth]{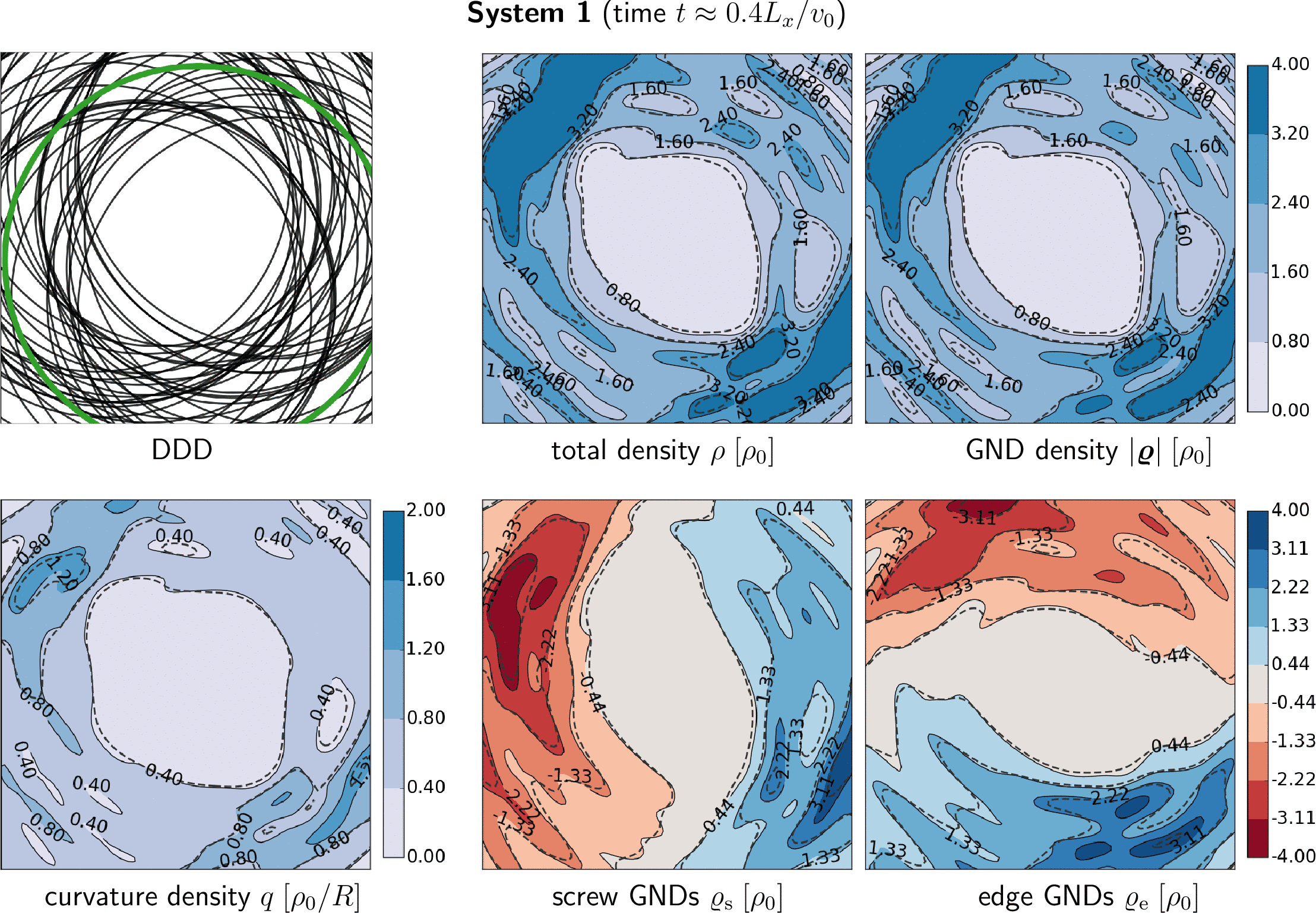}
	\caption{\label{fig:open} 	
		A snapshot in time (after segments traveled $\approx 0.4L_x$) for System 1 with open boundaries.  The color values and solid contour lines show reference DDD data, the dashed  lines show contours of the CDD data.}
\end{figure}

For this system a constant velocity of $v_0=0.01\,{\rm nm}/\upmu{\rm s}$ is prescribed, and time is expressed as multiple of the duration that density needs to completely traverse the domain, $L_x/v_0$. Boundaries are 'open', i.e. dislocations may leave the volume as if the domain were just a sub-domain of a much larger domain. 
This case of dislocation loops which have the same curvature and expand with constant velocity is a case for which the CDD theory is supposed to give mathematically exact results, and the system therefore can be used as a reference system for testing the quality of the numerical scheme and if the chosen standard deviation $s$ is appropriate. Initial values are shown in \figref{fig:inivals}. Two different simulations are run:  discrete microstructure is evolved with the DDD model and - separately - the continuous microstructure is evolved with the CDD model. To compare the two simulations at time $t$ the discrete microstructure was converted into CDD field quantities using the same strategy as for obtaining initial values from discrete data (in the sketch \figref{fig:sketch}F which is then compared with E).  \Figref{fig:open} shows a snapshot in time, where in the contour plots the dashed lines indicate CDD values and the full lines are the contours of the converted DDD values. 
We observe that as dislocation loops expand they flow away from the center region where the total density $\rhot$ tends to zero (compare $\rhot$ in \figref{fig:open}). The difference in the centers of the loops becomes more insignificant as their radii grow and the resulting density distribution takes a loop-like shape. In each point dislocations now have all approximately the same orientation, they become GNDs which shows in $\rhot\approx |\Bvarrho|$ as can be observed in \figref{fig:open}. The comparison of DDD and CDD fields shows  excellent agreement. Only minor features of the DDD fields are slightly smoothed out which is an artifact of the numerical solution.
\begin{figure}\centering
	\includegraphics[width=\textwidth]{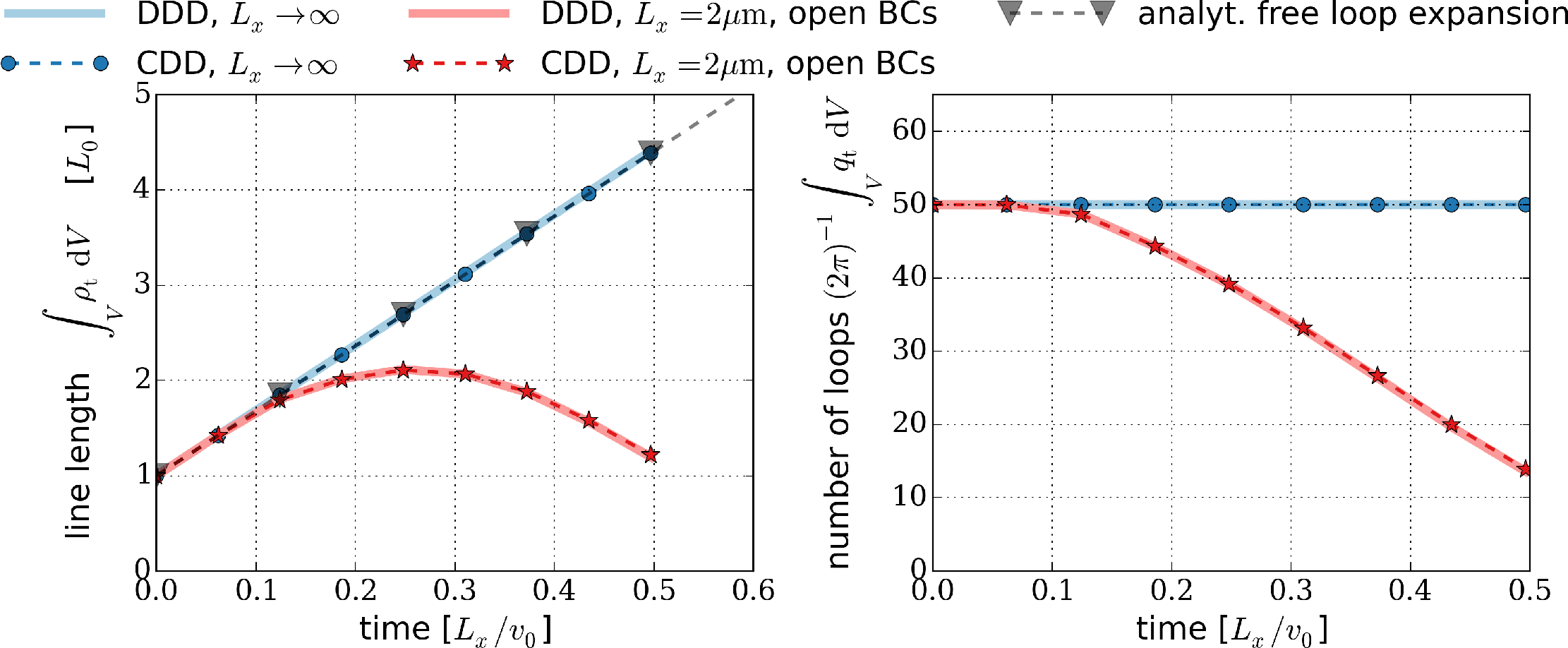}
	\caption{\label{fig:open:linelength} Normalized dislocation line length (with $L_0$ the initial line length) and total number of loops for System 1 as function of time. For comparison also the data for an unbounded system is shown together with the analytical solution.}
\end{figure}
The volume integrals of $\rhot$ and $\qt/2\pi$ give the total line length and the total number of dislocation loops and are shown in \figref{fig:open:linelength}. For reference we also show the line length for an infinite size system (approximated by a significantly larger domain in $x$ and $y$ direction) for which the total line length can be obtained analytically as $L(t)=N_{\rm d}2\pi(R+vt)$. Initially, both systems exhibit a linear line length increase (loops expand with constant velocity). As soon as density leaves the volume through the boundaries the line length for the finite size system decreases; the line length in the infinite system, however, continues to increase linearly. CDD data matches the DDD data perfectly and also coincides for the large system with the analytical solution. The right plot of \figref{fig:open:linelength} shows the total number of dislocation loops. This value stays constant for the infinite system, which is correct since there are neither sources nor annihilation. For the finite system the surfaces act as sinks and the number of loops is reduced. 
Note that this interplay between density and curvature density will also become relevant when sources and annihilation are to be modeled (which we do not consider within the present work).

\subsection{System 2: distribution of circular loops in a finite domain}
\label{sec:cell}
\begin{figure}[htb!]
\centering
\includegraphics[width=0.35\textwidth]{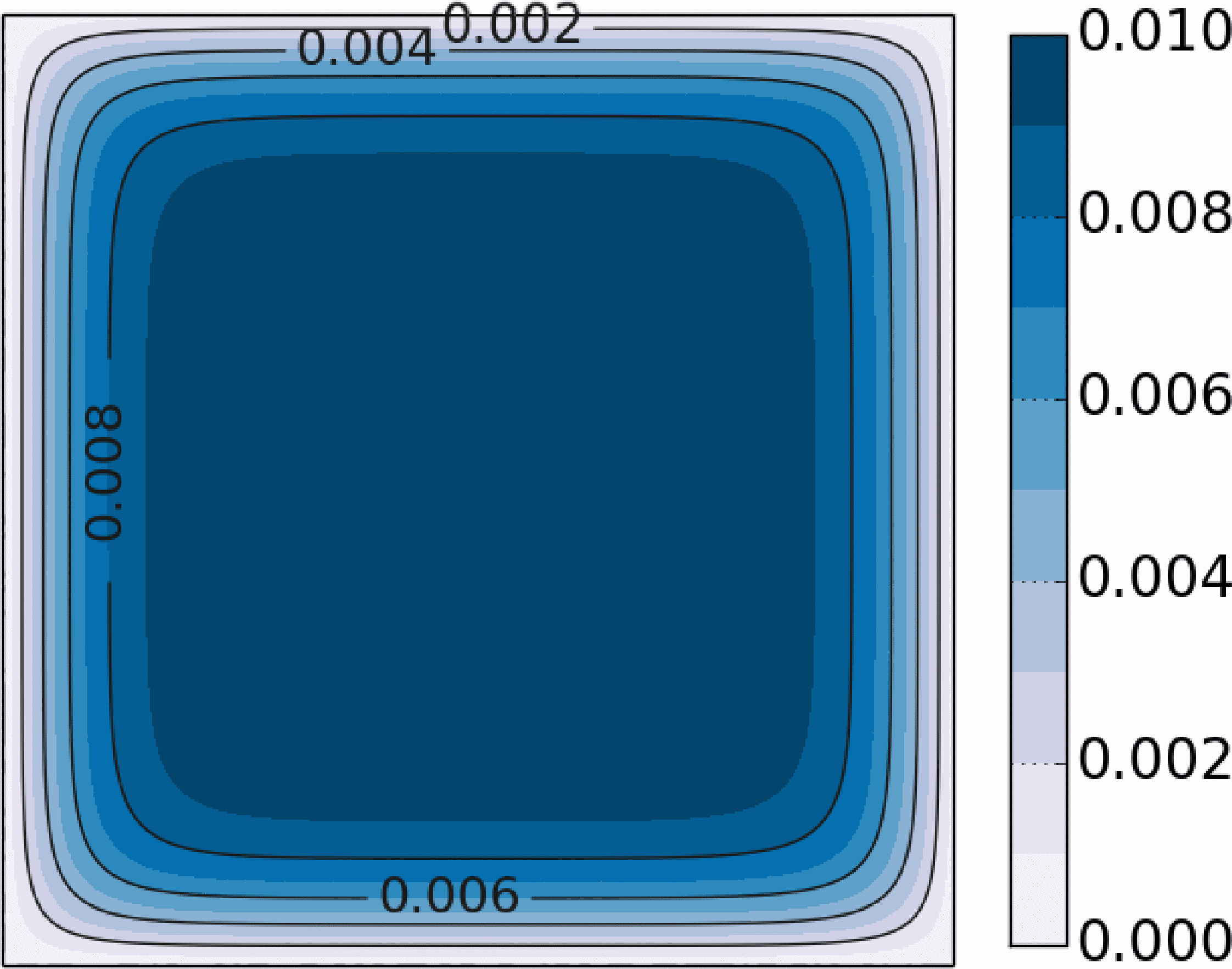}
\caption{\label{fig:cell:v}
Prescribed velocity field with $v_0=0.01\,\rm nm/\upmu s$ in the center region and $v=0$ at the boundaries of a 'grain' with impenetrable boundary conditions (System 2). This velocity field is used both for the CDD and the DDD simulation }
\end{figure}

\begin{figure}[htb]
\centering
%
\includegraphics[width=\textwidth]{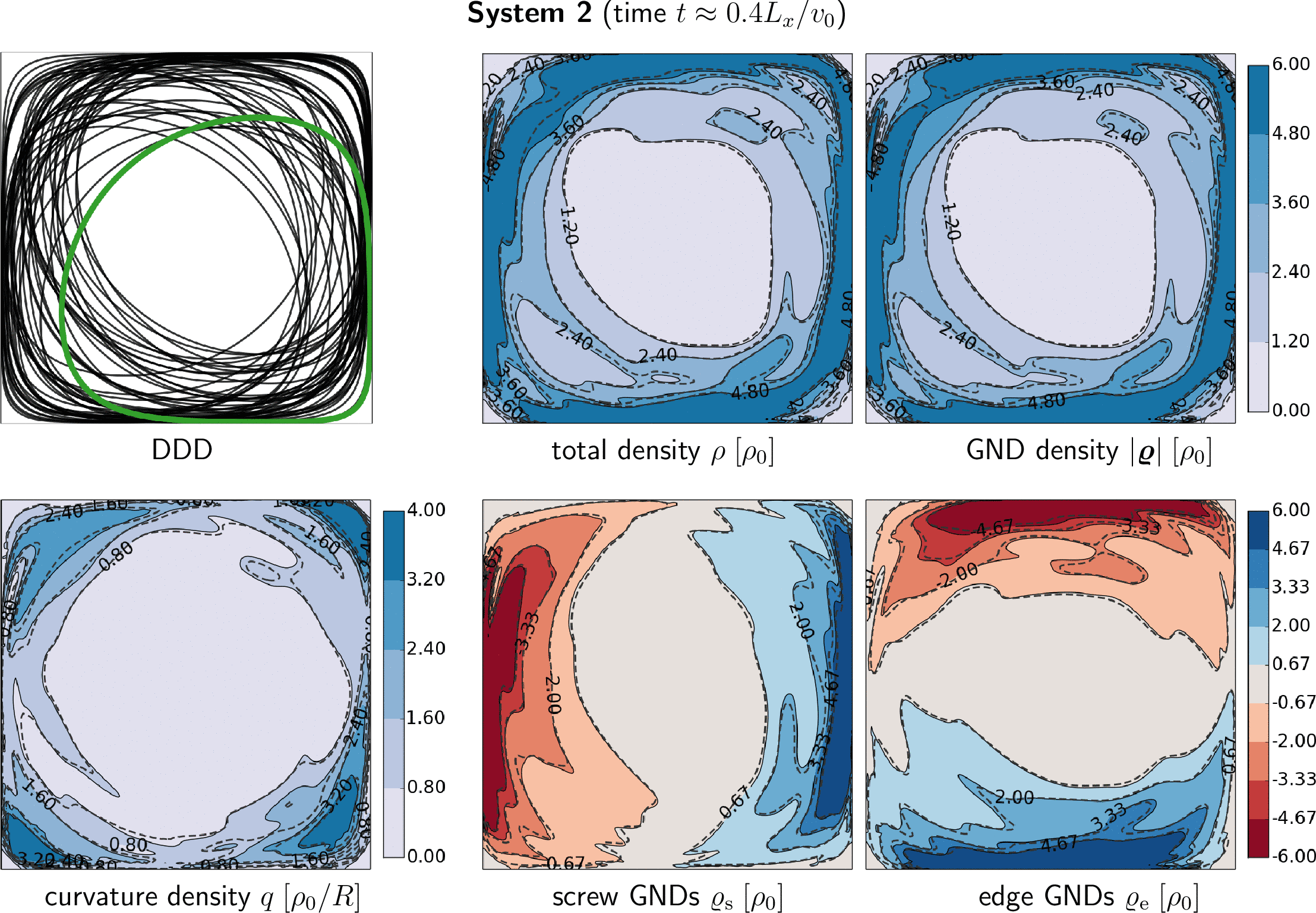}
\caption{\label{fig:cell:2D}
	Snapshot of dislocation structure in a quadratic grain with impenetrable boundaries. The solid contour lines show reference data obtained from DDD, the dashed lines show CDD data. }
\end{figure}

\begin{figure}[htb]
	\includegraphics[width=\textwidth]{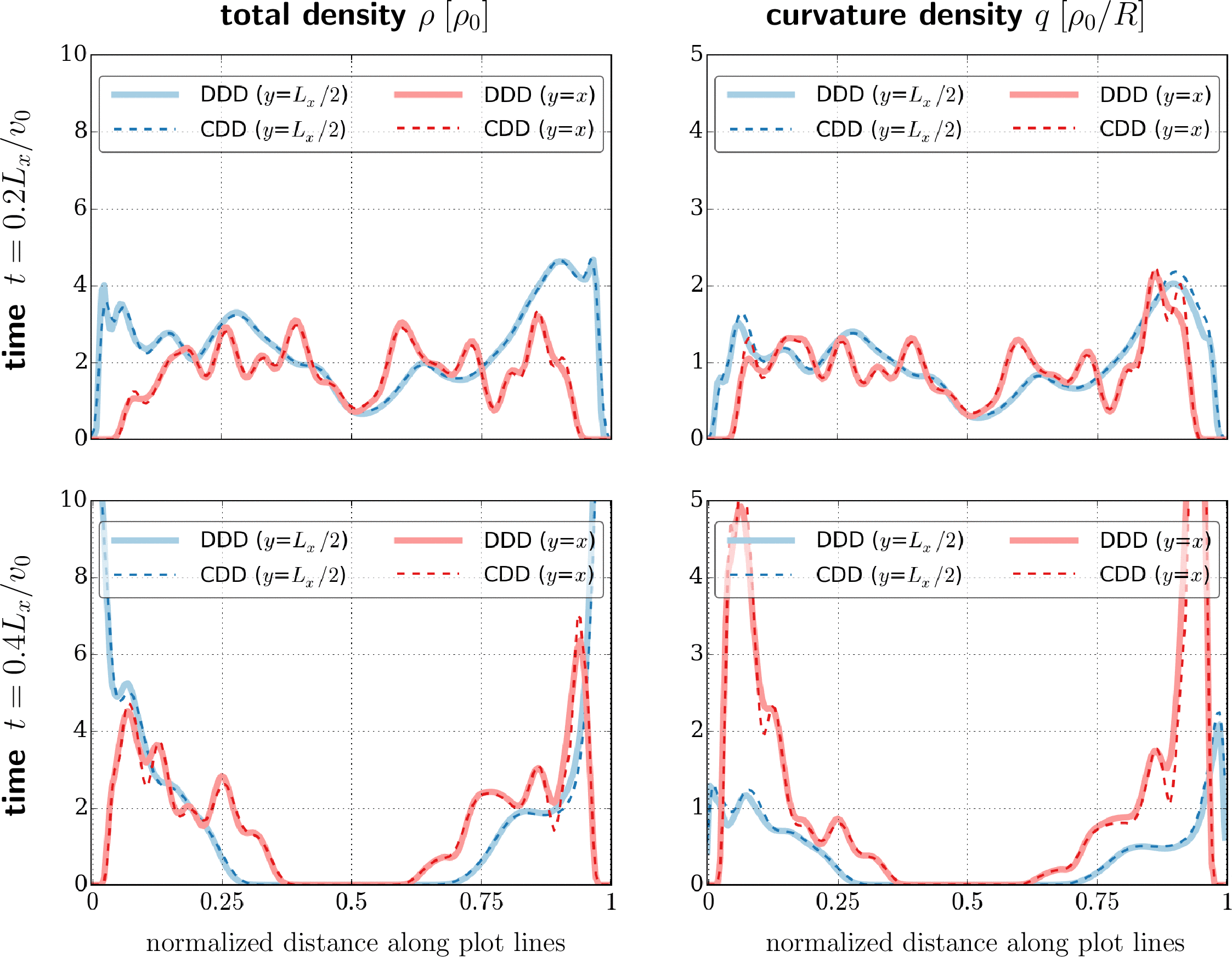}
	\caption{\label{fig:cell:1d}
		Snapshots in time for System 2: total density (left column) and curvature  density profiles (right column) for an early point of time (top row) and a later point of time (bottom row) which corresponds to \figref{fig:cell:2D}. The plotted data is along a horizontal line  at $y=L_y/2$ (blue) and a diagonal line at $y=x$ (red), where the lower left corner of the domain is at $x=y=0$ (cf. \figref{fig:sketch}).} 
\end{figure}

Now the outflow of dislocations will be constrained by imposing impenetrable boundary conditions, i.e.  the dislocation flux $\rho v$ is enforced to be zero at the boundaries. Numerically, this is done by prescribing a velocity field which decays smoothly to zero directly at the boundary (in physical terms this boundary layer could be imagined as the result of FIB damage in the surface-near regions). To mathematically model the boundary layer we use the following sigmoidal function, 
\begin{eqnarray}
 f(\xi)=\frac{2}{1+\exp[-\xi \beta]} - 1,
 \end{eqnarray}
where the parameter $\beta$ controls the shape of the function and is chosen such that the boundary layer is clearly visible and allows to study the dislocation flux in detail ($\beta=20$). $\xi$ is the normalized distance from the boundary, e.g. with $\xi=0\ldots 1$. The velocity field can now be composed from the four contributions
\begin{eqnarray}\label{eq:velocity_impBC}
	v(x,y)=v_0\cdot f(x/L_x)\cdot f(1-x/L_x)\cdot f(y/L_y)\cdot f(1-y/L_y).
\end{eqnarray}
It is depicted in \figref{fig:cell:v} and will be used for CDD as well as for DDD simulations.
Both simulations are conducted with the initial dislocation structure that was used as well for System 1 (\figref{fig:inivals}). Again, the DDD simulation starts from the discrete microstructure, the CDD simulation from the continuous microstructure as obtained by \emph{D2C} conversion. Both simulations evolve independently. Only in a post-processing step we again convert the resulting DDD microstructure for reference purposes into continuous field quantities as outlined before. A snapshot in time of the respective evolved dislocation field quantities is shown \figref{fig:cell:2D} with CDD values shown as dashed contour lines and converted DDD field values shown as solid lines  (also see the figure in the appendix and the supplementary movie 1 which additionally shows the evolution of average curvature $k=\qt/\rhot$).
During the simulation dislocation loops initially expand freely until parts of them reach the boundary layer within which the velocity decays to zero. The non-zero velocity gradient rotates line segments such that they eventually are parallel to the boundary, i.e. their line tangents are aligned with the contours of the velocity field. This causes the initial SSD density to increasingly become geometrically necessary when dislocations pile up against the boundaries (compare $\rhot$ and $|\Bvarrho|$ in \figref{fig:cell:2D}). Additionally, dislocations need to bend with a high curvature radius near the corners of the domain. This shows in the high value of the curvature density in these regions.
For a more detailed view \figref{fig:cell:1d} shows horizontal and diagonal line plots through the domain for two different time steps. We observe again an excellent agreement between the DDD and CDD field values for all times, and only small deviations can be seen close to the boundaries and in particular for the curvature density. 
%

\begin{figure}\centering
	\includegraphics[width=0.55\textwidth]{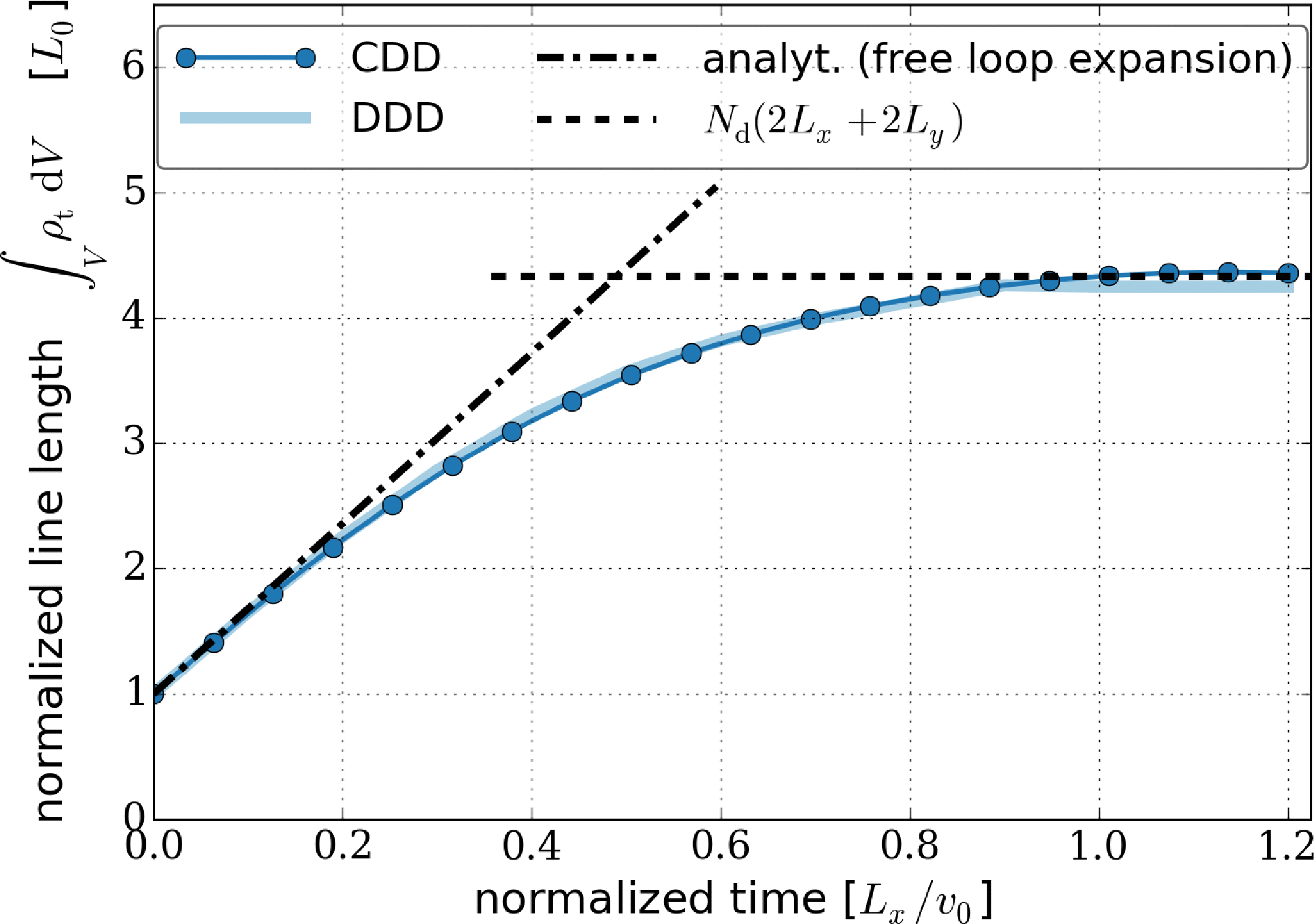}
	\caption{\label{fig:cell:linelength} Dislocation line length for System 2 as function of time. For comparison also the two analytical asymptotes are shown.}
\end{figure}

The total line length inside the system serves as a good plausibility test: initially we again would expect free loop expansion until finally all loops are aligned with the boundary and are (nearly) rectangular in shape. \Figref{fig:cell:linelength} shows the total line length vs time, and it is obvious that at all times DDD and CDD agree very well: both approach the horizontal tangent when all lines are aligned with the boundaries. The fact that DDD values are slightly below the horizontal tangent is correct because loops are not exactly rectangular but still have small rounded corners. For the DDD curve we directly used the line length from the discrete data while for the CDD data we first had to integrate $\rhot$. The small difference in the final value of the line length arises from discretization errors due to very steep gradients at the boundaries.  In this respect, more realistic systems that also consider elastic dislocation interactions  will numerically behave even better because pile ups always have a finite size.

\subsection{System 3: circular dislocation loops in a domain with an impenetrable inclusion}
\label{sec:inclusion}
\begin{figure}[htb]
	\includegraphics[width=\textwidth]{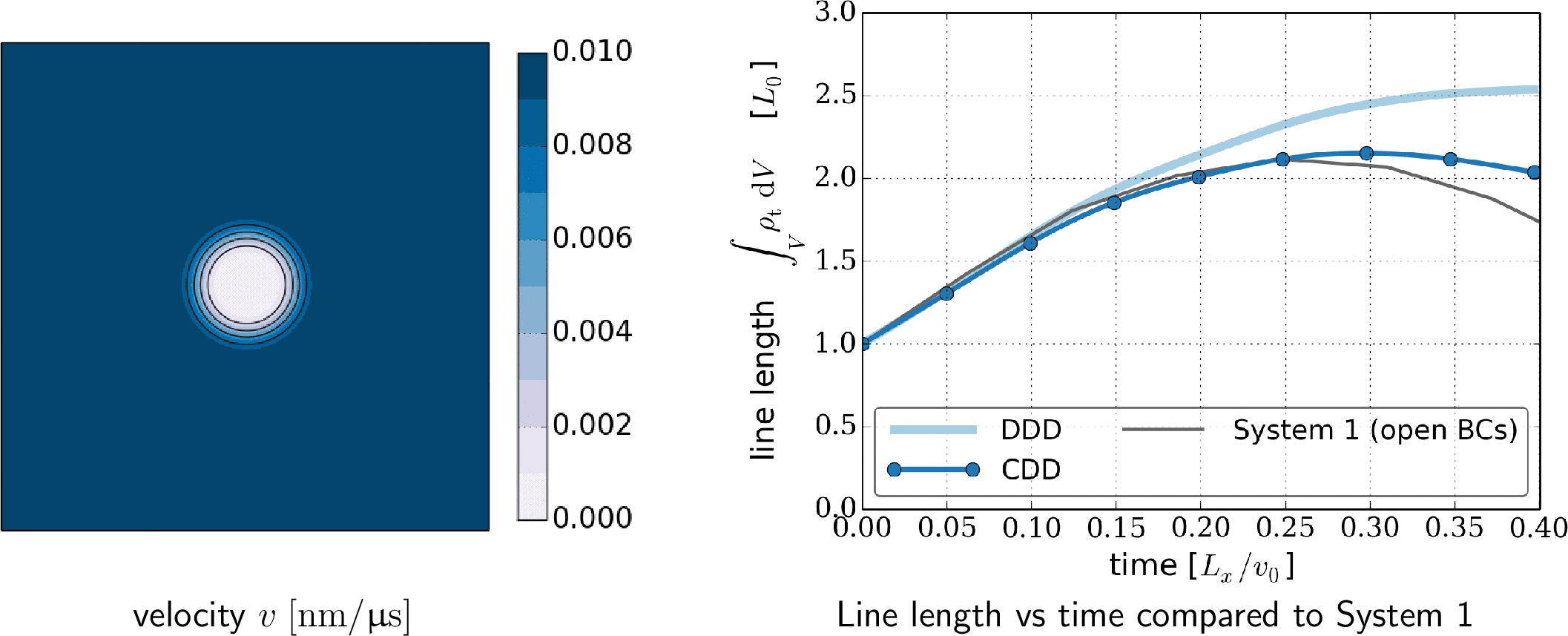}
	\caption{\label{fig:inclusion:vandL}
	Velocity field with the ``inclusion" in the center and evolution of line length (normalized with initial total line lenght $L_0$). }
\end{figure}

The third system studies the idealization of a plastically deforming matrix into which an elastic, circular inclusion is embedded. Dislocations can move freely in the matrix while the inclusion acts as an impenetrable obstacle for dislocation motion. Annihilation of lines is not considered, hence, the number of dislocation loops stays constant; for real systems this is a strong restriction, but again, the philosophy is to decouple interactions and reactions from the purely kinematic effects. 
The computational model is - as in System 1 - a quadratic domain ($L_x=L_y$) with open boundaries and contains the inclusion in the center. The inclusion itself is represented by an internal boundary which we model as a circular sub-domain (radius $R=L_x/10$) with zero velocity in the center region and a smooth transition from $v=0$ to a constant velocity $v_0$ away from the inclusion (cf. \figref{fig:inclusion:vandL}(a)). 
As a result of the velocity field dislocations are first slowed down on their way towards the inclusion and eventually freeze in their motion so that they cannot enter the inclusion. Both simulation methods use the same initial values as before. This does place some dislocations inside the inclusion which is unphysical. It turned out, however, that the influence on the resulting microstructure is only small. Therefore, we decided to consistently use the same initial values as for all other systems as well. Evolving the DDD and CDD system gives typical dislocation patterns as shown in \figref{fig:inclusion}, \Insert{additional plots including error plots can be found in the appendix.}
\begin{figure}[htb]
	\centering
	\includegraphics[width=\textwidth]{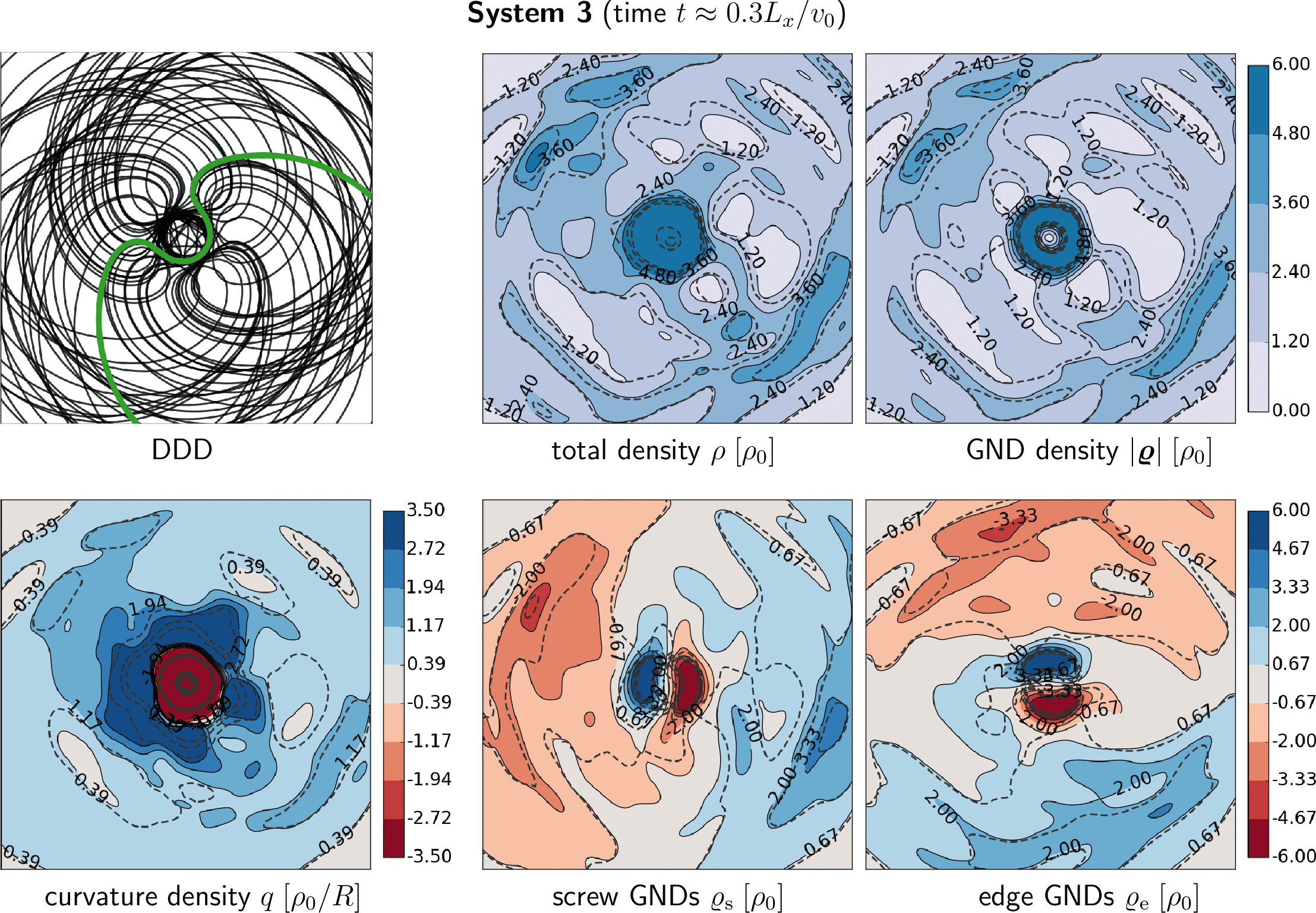}
	\caption{\label{fig:inclusion}Evolved dislocation loops in a domain with an precipitate, i.e. a region in the center with zero velocity. The color values and solid lines show reference DDD data, the dashed contour lines are CDD data. }
\end{figure}

\begin{figure}[htb]
	\centering
	\includegraphics[width=\textwidth]{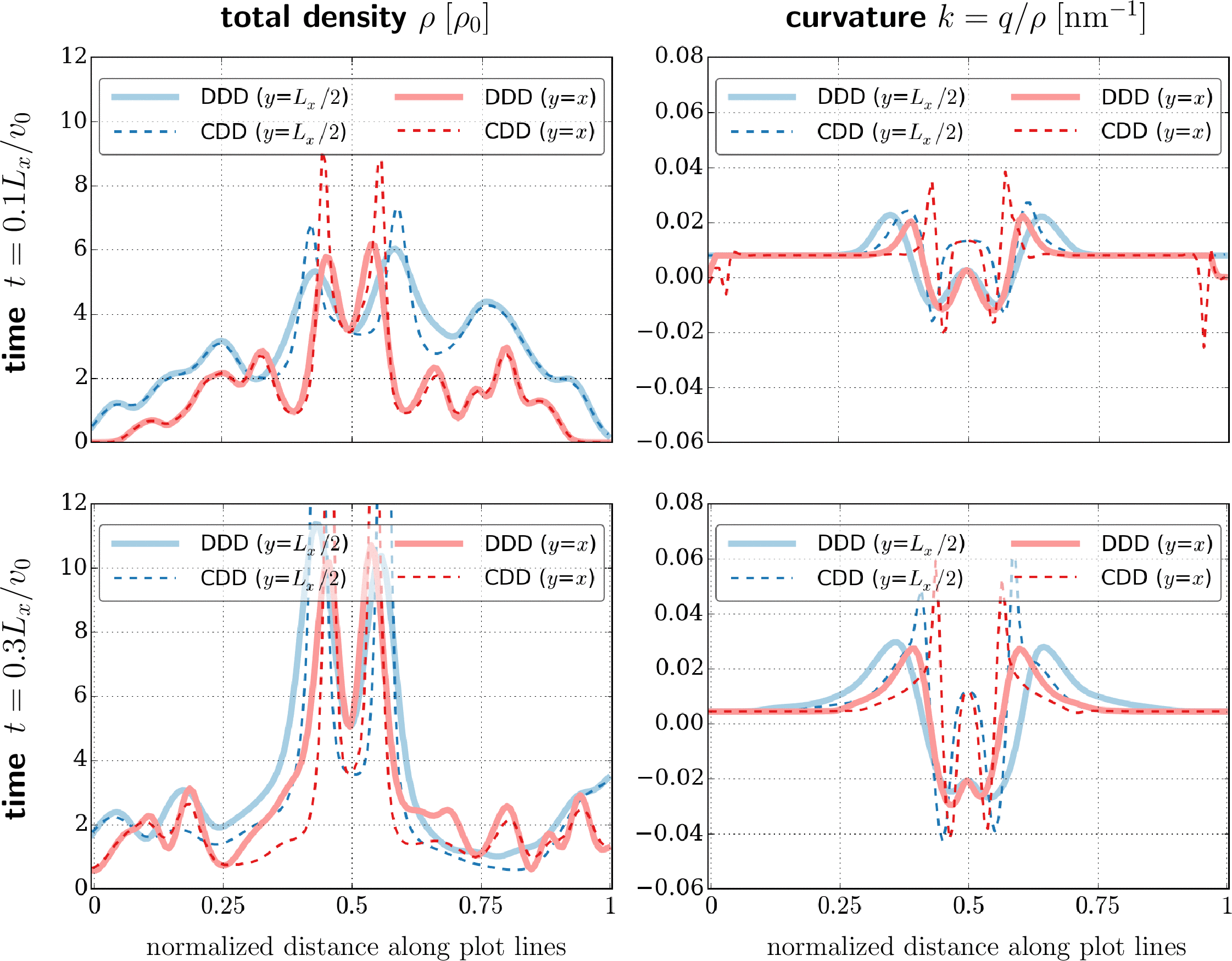}
	\caption{\label{fig:inclusion1d}
		Snapshots in time for System 3: total density (left column) and curvature  profiles (right column) for an early point of time (top row) and a later point of time (bottom row) which corresponds to \figref{fig:inclusion}. The plotted data is along a horizontal line  at $y=L_y/2$ (blue) and a diagonal line at $y=x$ (red), where the lower left corner of the domain is at $x=y=0$ (cf. \figref{fig:sketch}).
	}
\end{figure}

An auspicious feature is the formation of pile-ups of bent lines around the inclusion: when expanding dislocation loops touch the inclusion they get a 'dent' (see the colored line in \figref{fig:inclusion}(a)) and adjust to the shape of the inclusion. At the same time, line segments further away start 'spiraling' around the inclusion and deposit an increasing number of segments around the inclusion (also see the figure in the appendix and the supplementary movie). This behavior has a number of consequences and can be observed in the CDD representation as well:
\begin{itemize}
	\item \emph{Circular pile-ups of dislocation density} can be seen in the upper middle and right plots of \figref{fig:inclusion} : $\rhot$ has a high value in the vicinity of the inclusion because density from piled-up dislocations together with  pre-existing line segments in the inside of the precipitate add up. In conjunction with the GND density $|\Bvarrho|$ we infer that dislocations in the inside are mainly statistically stored while pile-ups around the inclusion are - literally - geometrically necessary.
	\item \emph{Circular GNDs around the inclusion form an inverted loop}, cf. \figref{fig:inclusion} bottom right and middle plot. Taking a look at the sign of e.g. the edge GNDs $\rhoe$ around the inclusion it is found that their orientation was rotated by $180^\circ$ (also compare the initial values in \figref{fig:inivals}). For instance the upper half is now negative and therefore would move downwards but is hindered by the inclusion. 
	\item \emph{Bent dislocations around the inclusion have negative curvature} - the above statement about an inverted loop already suggests that the curvature of dislocations around the inclusion should be negative which in fact can be observed in the two plots on the right of \figref{fig:inclusion1d} where the average curvature $k=\qt/\rhot$ was computed and plotted along two lines passing through the inclusion. The inclusion has a radius of $L_x/10$ which is equivalent to a curvature of $k_{\rm inc}=0.020\,{\rm nm}^{-1}$. Due to the smooth interface the velocity is zero only at about $L_x/16$ which is equivalent to a curvature of $k_{\rm min}=0.031\,{\rm nm}^{-1}$.  In the bottom right plot of \figref{fig:inclusion1d}  we find the negative curvature peak values at $|k_{\rm CDD}|\approx 0.040\, {\rm nm}^{-1}$ in good agreement with  both the geometry value $k_{\rm min}$ and the DDD curvature value of $|k_{\rm DDD}|\approx 0.028\,{\rm nm}^{-1}$.
	%
\end{itemize}

Deviations between DDD and CDD are larger than for the previous two benchmark models. They are particularly large in those regions which exhibit large gradients in the velocity and in the CDD field values, both of which cause increasing discretization errors. 
What is the average response of this system? Comparing the temporal evolution of the DDD line length in \figref{fig:inclusion} to System 1 shows that the inclusion creates additional density which is clearly visible from approximately $t=0.15 L_x/v_0$ on and is caused by lines bending around the inclusion. CDD is lagging behind and is initially even slightly below the line length of System 1 (due to the lower average velocity as compared to System 1). After  $t\geq 0.25 L_x/v_0$ the CDD model is then also showing an additional increase in line length. 

The difference between CDD and DDD  is the result of larger errors close to the inclusion where CDD almost always underestimates the true density values. Increasing the resolution of the spatial discretization of the finite element scheme does not yield any appreciable alleviation. The reason is the very complex deformation state of dislocations that brings the CDD theory to its limits because near the inclusion the line curvature can be different for different line orientations in the same averaging volume - a detail that cannot be represented with this simplified variant of CDD (this CDD theory only can represent \emph{one} average curvature value for each point/averaging volume). 
Although a number of details of this system have been predicted properly - e.g. the line curvature around the inclusion which is important when it comes to hardening effects  in terms of line tension - this particular CDD formulation only roughly predicts the line length production close to the inclusion. More elaborate evolution equations which contain more information about the orientation distribution and curvature of dislocations are required as e.g. those derived in \cite{Hochrainer2015_PhilMag, Monavari2014_MRSSP}. The \emph{D2C} strategy for more detailed CDD formulations, however, will still remain the same.

\SetTextColorRed
\subsection{\Insert{System 4: random dislocation distribution in a periodic domain}}
\label{sec:periodic}
\Insert{The previous benchmark systems have in common that they start with random initial values which during time evolution become strongly polarized through the geometrical constraints imposed by boundary conditions (e.g. dislocations adjust their line orientation and curvature to the shape of external or internal boundaries and become geometrically necessary). System 4  will now investigate how good CDD performs in a bulk-like situation where the total density consists to a large extend of statistically stored dislocation density superimposed with (smaller) GND density fluctuations. Periodic boundary conditions are used to eliminate geometrical constraints. Initial values consist of a statistically homogeneous  random distribution of 200 dislocation loops of the same radius, and their centers are distributed across the whole volume $L_x\times L_y\times \triangle z$. This results in the same average density $\rho_0$ as in the center region of Systems 1-3; all other parameters stay the same. }

\begin{figure}[htb]
	\centering
	\includegraphics[width=\textwidth]{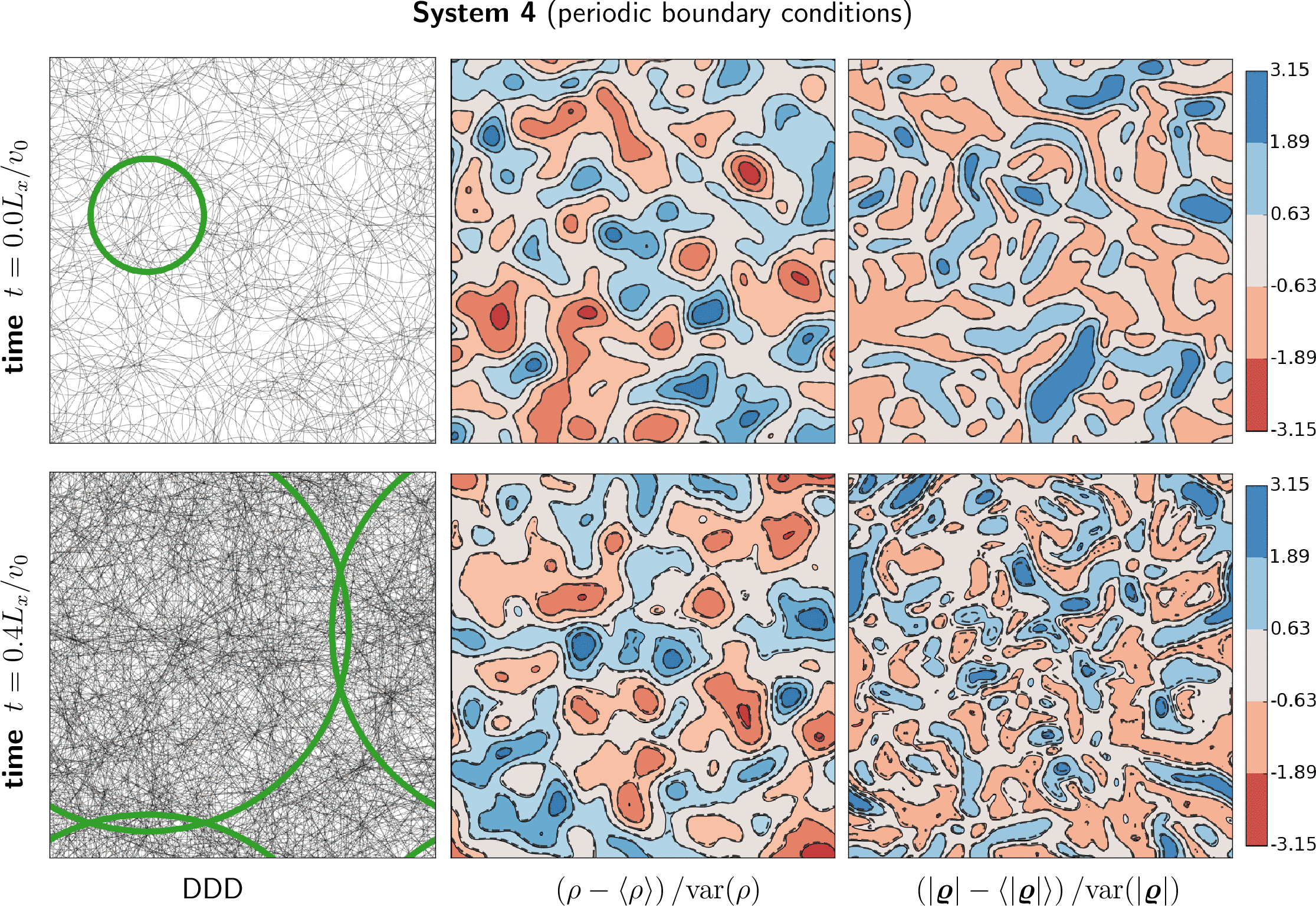}
	\caption{\label{fig:PBC2D} Initial values and evolved state of dislocation microstructure in a periodic domain. Contour plots show the normalized fluctuations of total density (middle column) and of the GND density (right column).  The color values and solid lines again denote reference DDD data, the dashed contour lines are CDD data. }
\end{figure}
\begin{figure}
	\includegraphics[width=0.95\textwidth]{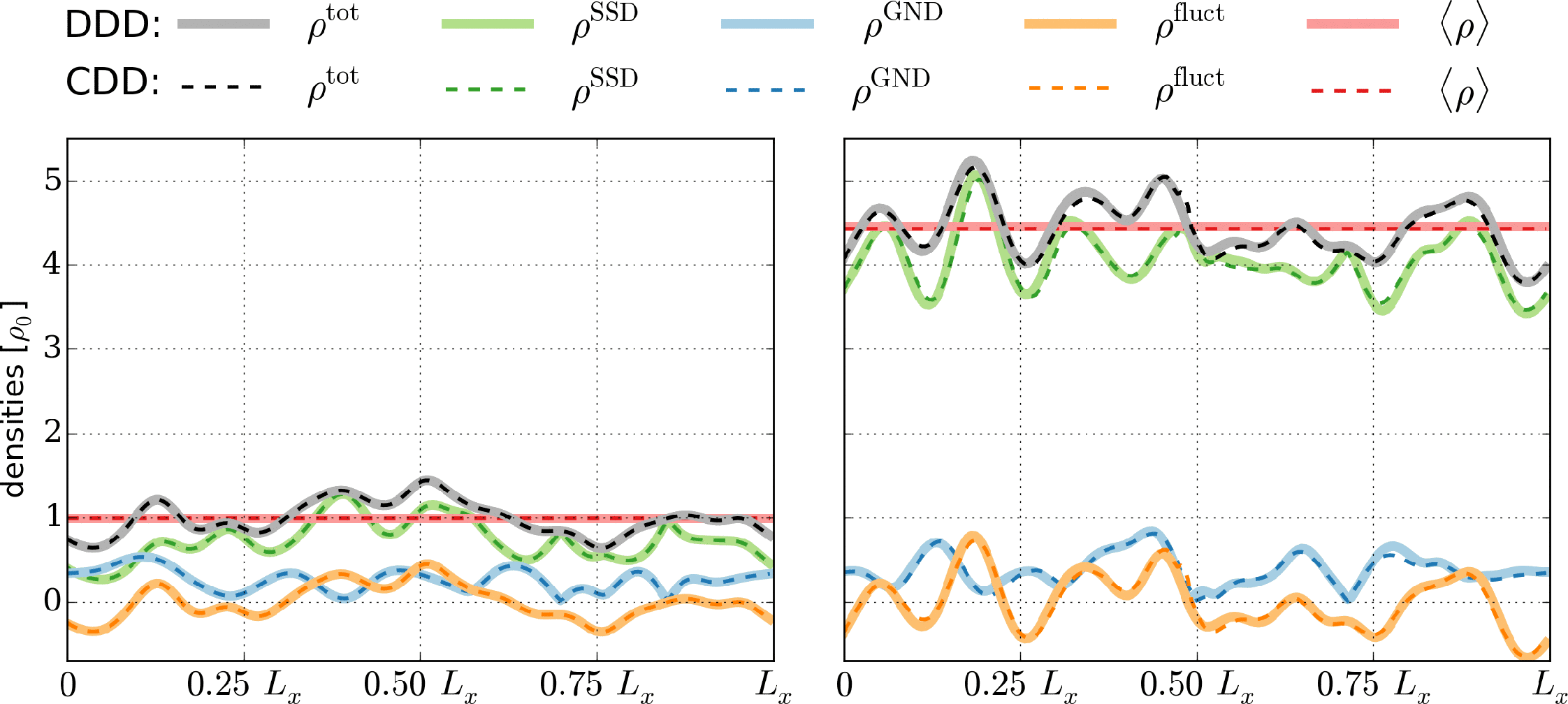}
	\caption{\label{fig:study4_lineplot}
		Densities along a horizontal line through the center of the computational domain for Study 4: initial state (left) and evolved state at $t=0.4L_x/v$ (right).
	}
\end{figure}

The velocity and loops' curvature radius are constant everywhere. Since this study analyzes the quality of the representation of flow and evolution of density fluctuations, it is useful to  decompose the total density into fluctuating and volume-averaged contributions, $\rhot(x,y) = \rhot_{\rm fluct}(x,y) + \langle\rhot\rangle$. For plotting the field data we additionally normalize the fluctuations with the variance of the respective field, e.g. in \figref{fig:PBC2D} we plot for the total density $(\rhot - \langle\rhot\rangle)/{{\rm var}(\rhot)}$.
%
\Figref{fig:study4_lineplot} shows line plots of initial and evolved DDD and CDD data, additional error plots can be found in the appendix. Clearly, all details of the discrete microstructure are accurately reproduced. Fluctuations evolve in a non-trivial manner and we attribute negligible deviations to numerical errors. The average behavior matches exactly the DDD values. 
The overall very good agreement is not surprising after seeing that already System 1, which initially had a similar random structure in the center of the domain, gave very good agreement between DDD and CDD. It is, however, important to show that this good agreement is sustained during time evolution  and that even the increase in the range of the fluctuations are properly represented. An application based on this type of system has been studied in the context of dislocation patterning \cite{Sandfeld2015_MSMSE} where in particular the fluctuations together with the correct line length increase and a Taylor-type flow stress were identified as key ingredients for pattern formation.


\SetTextColorBlack

\section{Determining the kinematic closure parameter $\Phi$ from DDD}
\label{sec:fitting}

The numerical studies in \secref{sec:examples} demonstrated that DDD simulations can serve as an elegant means for directly benchmarking continuum dislocation microstructure and its evolution. Thus, it seems natural to directly use DDD simulations to verify or even obtain expressions that were used to constitutively close the set of CDD evolution equations. Within the CDD equations occurred in particular one quantity which had to be assumed: the interpolation function $\Phi$, which was taken  in the 'maximum entropy approximation' approach as $\Phi=(|\Bvarrho|/\rhot)^2 (1+(|\Bvarrho|/\rhot)^4)/2$. $\Phi$ was in an earlier approach assumed to interpolate linearly between SSD and GND density. We will now analyze DDD simulations in order to decide which assumption is under realistic simulation conditions the most suitable one. Recall the definition of the evolution equation of the curvature density and $\BA^{(2)}$:
\begin{eqnarray}
\label{dqdt_inv}
\partial_t\qt^{\rm CD} &=&-\div( -v\BQ^{(1)} + \BA^{(2)}\cdot \nabla v ),\\
\label{A2_inv}
{\BA}^{(2)} &=& \frac{\rhot}{2}\left[ (1+\Phi) {\Bl}_{\varrho} \otimes {\Bl}_{\varrho} + (1-\Phi){\Bl}_{\varrho}^{\perp} \otimes {\Bl}_{\varrho}^{\perp}\right] . 
\end{eqnarray}

\begin{figure}[tbph]
	\centering
	\includegraphics[width=0.5\textwidth]{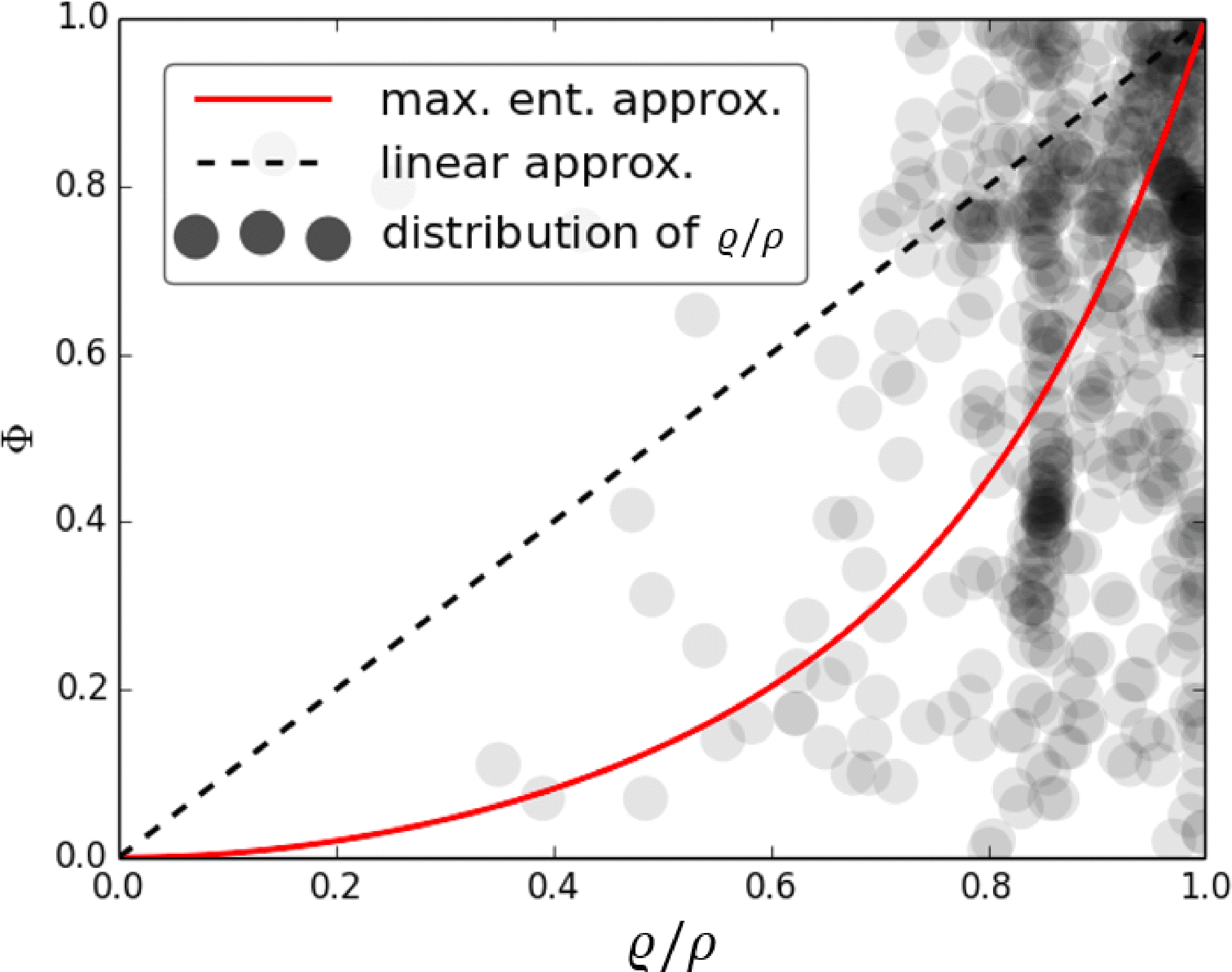}
	\caption{\label{fig:Phi}
		$\Phi(\varrho/\rhot)$ for the linear approximation and the 'maximum entropy approximation'. Data points show best fits of the numerically computed values for $\Phi$.}
\end{figure}

In $\BA^{(2)}$ the approximation $\Phi$  occurs in a vector product together with a spatial velocity gradient. Thus, $\partial_t\qt^{\rm CD}$ is sensitive w.r.t. variations in $\Phi$ only in regions where the velocity field is non-constant, as e.g. in the boundary layer close to the impenetrable boundary of System 2  from \secref{sec:cell}. Continuous CDD field variables can be obtained from a DDD simulation as outlined before, and the time derivative $\partial_t\qt$ can be numerically approximated by an explicit Euler scheme in time,
\begin{eqnarray}\label{eq:dq_DD}
\partial_t\qt^{\rm DD} = \frac{\qt^{\rm DD}(t_{i+1})-\qt^{\rm DD}(t_i)}{\Delta t} .
\end{eqnarray} 
At the same time also the evolution equation for $q$ (eqn. \eqref{dqdt_inv}) can be evaluated based on field data obtained from DDD ($\BQ^{(1)}$ and $\nabla v$ are both known, and $\Bl$, $\Bl^\perp$ are functions of $\Bvarrho$ and $\rhot$)%
\footnote{We note, that the space dependent tensors $\BQ^{(1)}$ and $\BA^{(2)}$ also can be  extracted from DDD data. E.g. the '11' component of $\BA^{(2)}$ can be obtained by integrating the discrete density for each separate line segment against the $\cos^2\varphi_\Bc$ of the discrete line orientation $\varphi_\Bc$, followed by regularization based on the Gaussian convolution.}. The unknown value of $\Phi$ can now be obtained for each point (where $\nabla v\neq 0$) from solution of the inverse problem :
\begin{eqnarray}
\textrm{\bf \textbf{Find}}\; \Phi(x,y) \; \textbf{-- for all}\; (x,y) \; \textbf{with}\; \nabla v(x,y)\neq 0 \; \textbf{-- such that} \;\nonumber\\
\qquad\|\partial_t\qt^{\rm DD}(x,y)-\partial_t\qt^{\rm CD}(\Phi(x,y); x,y)\|\rightarrow{\rm MIN}. \nonumber 
\end{eqnarray}
Note that no CDD simulation needs to be done, only DDD data is used. For each point $(x,y)$ we record the optimum value of $\Phi$ together with the local total density $\rhot(x,y)$ and GND density $\varrho(x,y)$. Plotting these data gives the distribution shown in \figref{fig:Phi}. There, the point of time was chosen such that already a significant amount of density reached the boundary layer but did not yet reach a stationary state. For an earlier point of time dislocations would not yet have reached the region where $\nabla v\neq 0$ and no data could be obtained. At a later point of time most dislocations form pile-ups of geometrically necessary configurations near the boundaries which results in a strong accumulation of data points at the top right corner of the diagram. 
It is obvious that the optimization strategy yielded points with a relatively large scatter. However, it can be concluded that most data points are located \emph{below} the straight dashed line which is the linear interpolation between GNDs and SSDs. The polynomial maximum entropy approximation fits much better. Binning the data e.g. in 20 $\varrho / \rho$ intervals gives averages that for $\varrho/\rho=0.8..1.0$ coincide nicely with the polynomial approximation. Nonetheless, the data is not sufficient for any statistical analysis in all other regions and could not be much improved through additional simulations.
These results might indicate that the approximation of $\partial_t q$ is not sufficiently sensitive w.r.t. to the only parameter $\varrho/\rhot$. Currently, higher order closure assumptions are under investigations. Their results might help to understand the difficulties in identifying a unique functional $\Phi$ for closure. 

\section{Conclusion}

We developed  a systematic method for averaging geometrical properties of discrete dislocation lines which allows for a direct comparison of discrete and continuum  descriptions of evolving dislocation microstructures. Particular care was taken to formulate the numerical approximation such that discrete and continuous formulations are at all times consistent with each other (e.g. both the DDD and the CDD descriptions were based on the same spline representation). 
Using DDD data as reference the comparison of simulation results for 4 different systems revealed a surprising accuracy of the CDD theory and its numerical implementation: CDD is able to evolve very complex dislocation microstructure and simultaneously respect various physical boundary conditions in almost exactly the same level of detail as DDD simulations do. Even in situations where the simplifying assumptions of the used CDD formulation were clearly violated (System 3) we found that results were still reasonably. 
Within this work we only considered the kinematic closure and assumed a (fixed and analytical) velocity function. An important step for future work which still requires a large amount of fundamental work, however, is the ``dynamical closure", i.e. the question what the average velocity of \emph{interacting} continuum distributions of dislocations under stress is.

Considering that in 3D DDD simulations the computational cost scales approximately with $N^2$ ($N$ being the number of nodes/segments) and considering that for a continuum theory the computational cost does not increase with the number of dislocations at all (not even when dislocation interactions are considered) we are very optimistic that a continuum description of dislocation plasticity might very soon complement DDD simulations in particular when it comes to high densities in large volumes and/or high accumulated plastic strains. 
Furthermore, the \emph{D2C} conversion might even be directly beneficial for analyzing DDD data: the continuum representation is well suitable for ensemble averaging and could be a novel way of directly comparing discrete microstructures.

\section*{Acknowledgments}
Financial support from the Deutsche Forschungsgemeinschaft (DFG) through Research Unit FOR1650 'Dislocation-based Plasticity' (DFG grants SA2292/1-1 and SA2292/1-2) is gratefully acknowledged.

	\setcounter{section}{0}
	\setcounter{subsection}{0}
	\setcounter{figure}{0}
	\setcounter{table}{0}
	\renewcommand\thesection{Appendix}
	\renewcommand\thefigure{\Alph{section}\arabic{figure}}
	\renewcommand\thetable{\Alph{section}\arabic{table}}

\setcounter{equation}{0}
\renewcommand{\theequation}{A\arabic{equation}}

\section{On the resolution of continuum fields}
\SetTextColorRed
In \secref{sec:evoeqns} continuous fields were obtained that had to be artificially smoothed out for numerical reasons. This was done by convolving the coarse grained fields (eqns. \eqref{eq:rho_i}-\eqref{eq:q_i}) with a discrete Gauss function $G^{\rm d}$. Choosing the standard deviation $s$ in \eqref{eq:rho_gauss} is a choice which has to be made for every system and e.g. with regards to numerical aspects as well as based on the desired degree of detail which the density fields should be able to represent (e.g. in terms of density fluctuations). In the following, we  summarize some key aspects that one might want to consider for deciding on a suitable value for $s$.

\begin{figure}[htb]
	\centering
	\includegraphics[width=\textwidth]{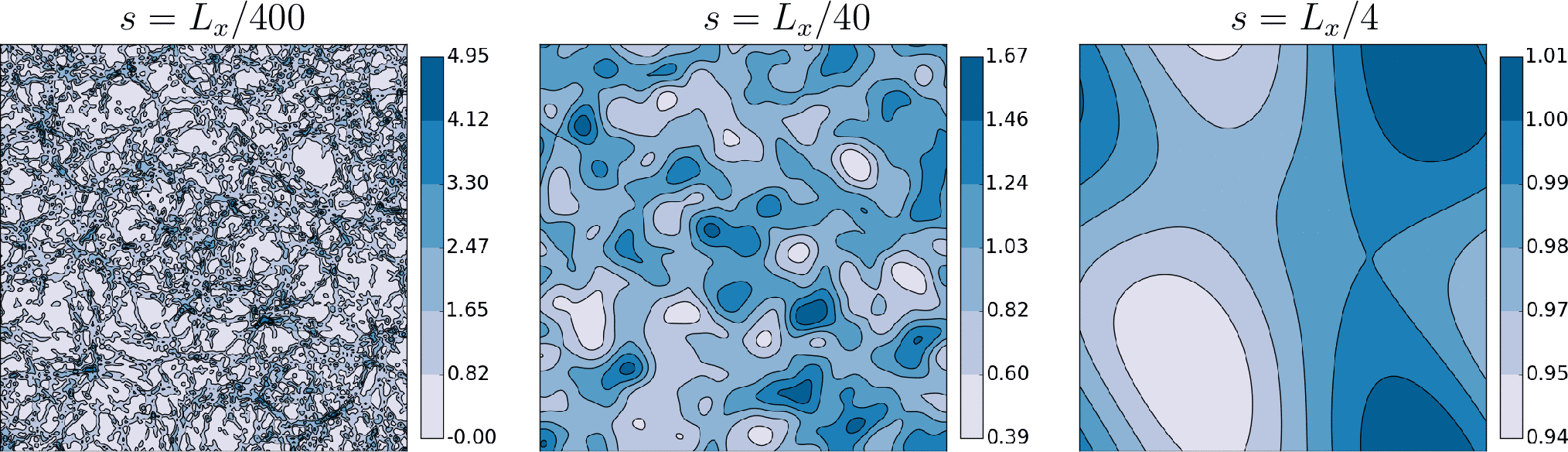}
	\caption{\label{fig:sigma:app} Initial density $\rhot$ for different values of the standard deviation $s$ obtained from the DDD data of System 4. The left plot is a nearly discrete microstructure, while in the right plot an almost homogeneous structure without any gradients resulted. CDD can in principle evolve all initial values accurately, the only limitation is our used numerical scheme, which does not work for values of below $s\approx L_x/200$.}
\end{figure}

\begin{itemize}
	\item The obvious criterion for a \textbf{lower bound} for $s$ is related to the resolution of the numerical scheme: choosing $s$ much smaller than roughly the size of a finite element (which additionally depends on the used shape functions) results in discretization errors, i.e. the fluctuations of the density field cannot be numerically represented properly anymore. In principle the discretization error can be measured. For practical purposes one could e.g. compare the numerical representation of a 'continuum loop' with the analytical solution. Note that choosing a value of the standard deviation in the range of a Burgers vector results in a computationally very expansive continuum model of nearly discrete objects (\figref{fig:sigma:app} left).
	\item A numerically stable solution of the transport equations of density fluxes additionally requires a certain degree of smoothness in order to avoid strong, undesired numerical oscillations. Studying the aspect of numerical stability is non-trivial (see e.g. \cite{Varadhan:2006vv} for a stability analysis for a continuum dislocation dynamics model based on the Kr\"oner-Nye tensor) and was not attempted for the present CDD equations so far. We approach this problem in a pragmatic way and choose numerical parameters as well as the \textbf{lower bound} for the standard deviation $s$ such that a single continuum dislocation loop can be expanded accurately within a small error tolerance during the simulated time \cite{Sandfeld2010_PhilMag90}.
	\item Deciding on an \textbf{upper bound} for the standard deviation $s$ is problem specific: CDD works properly for \emph{all} values of $s$ exceeding the above introduced lower bounds. However, the larger the chosen $s$ the more details of the heterogeneous microstructure will be destroyed. In particular, choosing $s$ in the range of magnitude of the system dimension causes all gradients of the density fields to vanish. As a consequence, the partial differential equations for transport of densities deteriorate to ordinary differential equations, which are no longer able to describe fluxes on a scale below the system size (\figref{fig:sigma:app} right).
\end{itemize}

In this work, the emphasis is on heterogeneous dislocation microstructure. Hence, we chose, guided by the mean dislocation spacing of $s\approx L_x/35$, a value of $s$ which is small enough that fluctuations are not getting smeared out, but which at the same time is large enough so that discrete dislocations can no longer be differentiated. 

\SetTextColorBlack

\section{Additional time evolution and error plots for System 2, 3 and 4}\label{sec:app:plots}
\begin{figure}[htb]
	\centering
	\includegraphics[width=\textwidth]{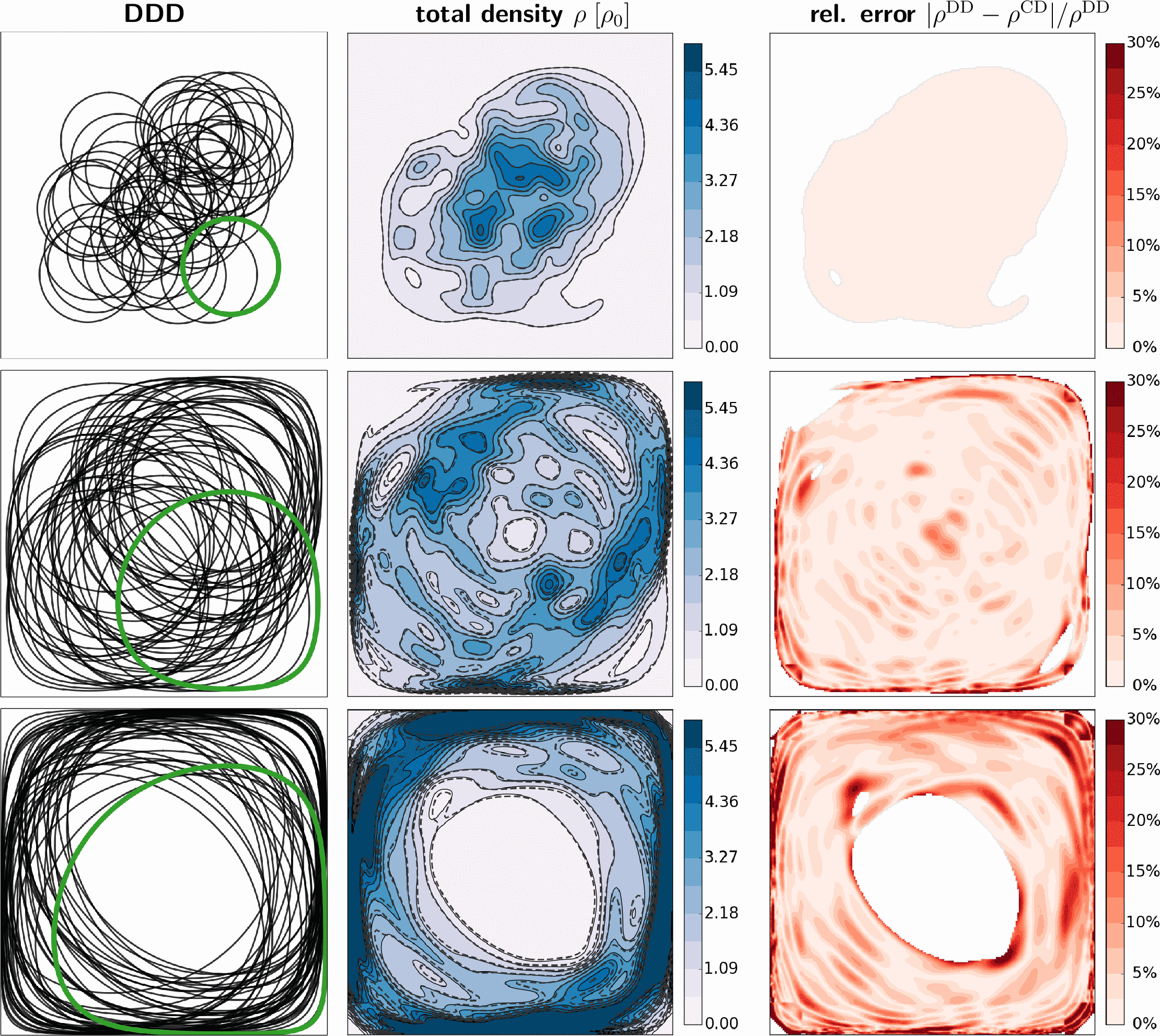}
	\caption{\label{fig:cell:app} System 2 - DDD data, density and relative error at time steps $t=0,\, 0.2$ and $0.4 L_x/v_0$ (from top to bottom). The white regions in the error plots indicate that the relative error was not computed due to (nearly) zero density.}
\end{figure}

\begin{figure}[htb]
	\centering
	\includegraphics[width=\textwidth]{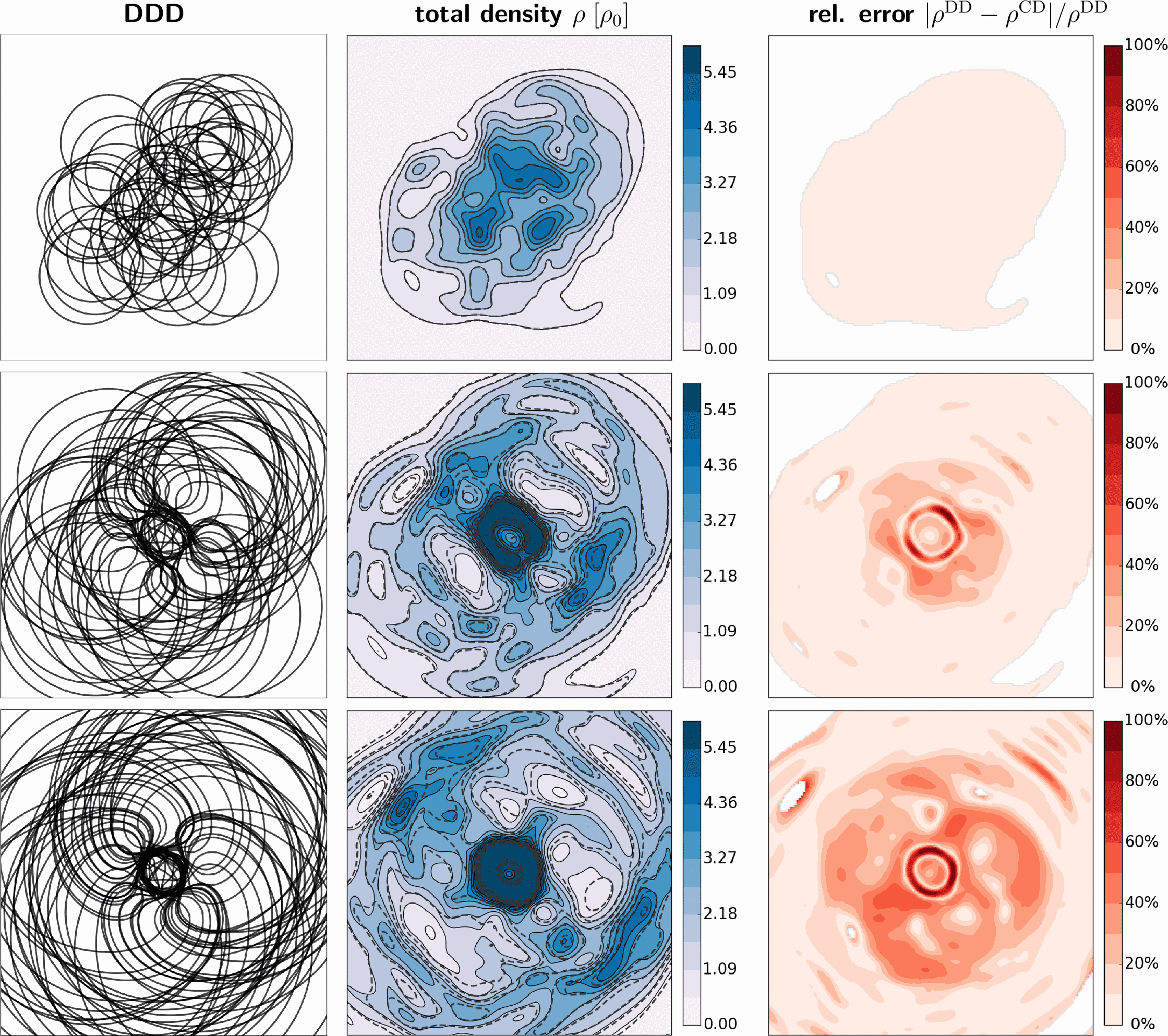}
	\caption{\label{fig:inclusion:app} System 3 - DDD data, density and relative error at time steps 
	$t=0,\, 0.135$ and $0.27 L_x/v_0$ (from top to bottom). The white regions in the error plots indicate that the relative error was not computed due to (nearly) zero density.}
\end{figure}

\begin{figure}[htb]
	\centering
	\includegraphics[width=\textwidth]{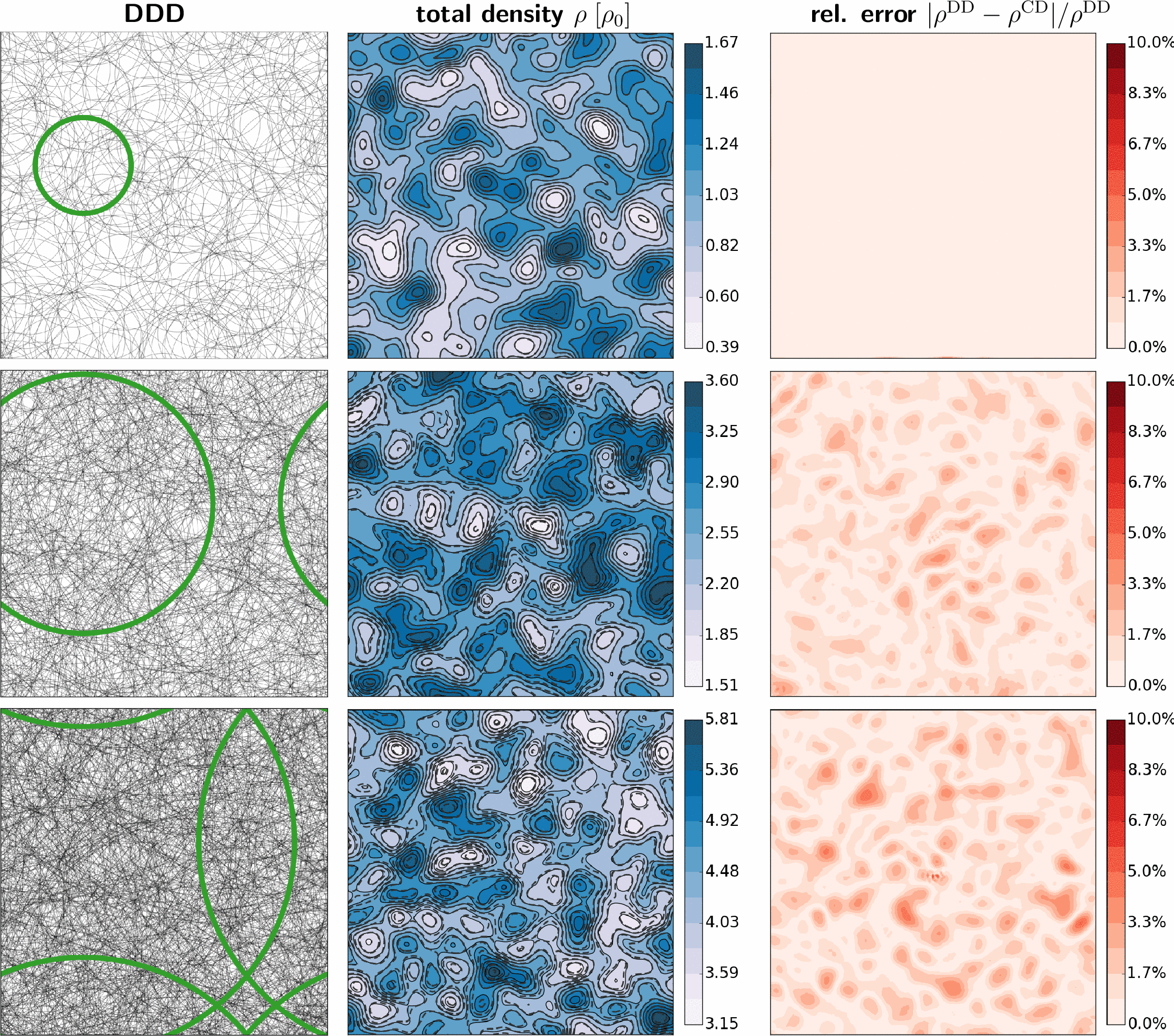}
	\caption{\label{fig:pbc:app} System 4 - DDD data, density and relative error at time steps $t=0,\,0.25$ and $0.5 L_x/v_0$ (from top to bottom). Note the changing color range for the total density. }
\end{figure}

\SetTextColorRed
In \figref{fig:cell:app} - \ref{fig:pbc:app} the time evolution of discrete and continuous microstructure together with the relative error $|\rho^{\rm DD}-\rho^{\rm CD}|/\rho^{\rm DD}$ for System 2 (grain with impenetrable boundaries), System 3 ('precipitate') and System 4 (periodic BCs) are shown. In order to reduce the unreasonable diverging behavior of the relative error  in regions where $\rho^{\rm DD}\rightarrow 0$ we chose a cut off value of 0.5 (i.e. less than half the size of the smallest interval of the color scale of the density contour plot for System 2 and 3) below which the error is not computed (shown as white regions). In the density plots the solid lines are -- as before -- the contours of the converted DDD data while the dashed lines show the CDD data.

\emph{System 2} in \figref{fig:cell:app} has in most regions and for all time steps relative errors of less than $\approx 5\%$. Only at later time steps when dislocations pile up against the boundaries and density exhibits steep gradients deviations become larger in some places. The CDD theory does represent the evolution of this system properly.
\emph{System 3} in \figref{fig:inclusion:app} shows clearly that the geometrical constraint of the inclusion creates a complex situation which only partially can be represented by CDD and causes large errors. Despite the fact that a number of features are correctly represented (in particular the curvature close to the inclusion, see main text), a more precise evolution of this system would require a more refined CDD theory.
\emph{System 4} in \figref{fig:pbc:app} demonstrates that CDD consistently and accurately predicts the increase of average density as well as the transport of density fluctuations - both of which are automatically contained within the set of evolution equations. The relative error is in most places even smaller than 4\%.


%

\SetTextColorBlack

\section*{References}

\bibliographystyle{unsrt}
\bibliography{paper}

\end{document}